\documentclass[twocolumn,tighten]{aastex631} 

\usepackage{apjfonts}
\usepackage{amsmath}
\usepackage{physics}
\usepackage{booktabs,array,multirow}

\usepackage{amssymb}
\usepackage{color}
\usepackage{mathtools}
\usepackage{ulem}
\usepackage[all]{hypcap}

\graphicspath{{./}{figures/}}

\begin{document}

\title{Stellar Escape from Globular Clusters. I. Escape Mechanisms and Properties at Ejection}

\author[0000-0002-9660-9085]{Newlin C. Weatherford}
\affil{Department of Physics \& Astronomy, Northwestern University, Evanston, IL 60208, USA}
\affil{Center for Interdisciplinary Exploration \& Research in Astrophysics (CIERA), Northwestern University, Evanston, IL 60208, USA}

\author[0000-0003-4412-2176]{Fulya K{\i}ro\u{g}lu}
\affil{Department of Physics \& Astronomy, Northwestern University, Evanston, IL 60208, USA}
\affil{Center for Interdisciplinary Exploration \& Research in Astrophysics (CIERA), Northwestern University, Evanston, IL 60208, USA}

\author[0000-0002-7330-027X]{Giacomo Fragione}
\affil{Department of Physics \& Astronomy, Northwestern University, Evanston, IL 60208, USA}
\affil{Center for Interdisciplinary Exploration \& Research in Astrophysics (CIERA), Northwestern University, Evanston, IL 60208, USA}

\author[0000-0002-3680-2684]{Sourav Chatterjee}
\affil{Tata Institute of Fundamental Research, Homi Bhabha Road, Mumbai 400005, India}

\author[0000-0002-4086-3180]{Kyle Kremer}
\affil{TAPIR, California Institute of Technology, Pasadena, CA 91125, USA}
\affil{The Observatories of the Carnegie Institution for Science, Pasadena, CA 91101, USA}

\author[0000-0002-7132-418X]{Frederic A. Rasio}
\affil{Department of Physics \& Astronomy, Northwestern University, Evanston, IL 60208, USA}
\affil{Center for Interdisciplinary Exploration \& Research in Astrophysics (CIERA), Northwestern University, Evanston, IL 60208, USA}

\begin{abstract}
The theory of stellar escape from globular clusters (GCs) dates back nearly a century, especially the gradual evaporation of GCs via two-body relaxation coupled with external tides. More violent ejection can also occur via strong gravitational scattering, supernovae, gravitational wave-driven mergers, tidal disruption events, and physical collisions, but comprehensive study of the many escape mechanisms has been limited. Recent exquisite kinematic data from the Gaia space telescope has revealed numerous stellar streams in the Milky Way (MW) and traced the origin of many to specific MWGCs, highlighting the need for further examination of stellar escape from these clusters. In this study, the first of a series, we lay the groundwork for detailed follow-up comparisons between Cluster Monte Carlo (\texttt{CMC}) GC models and the latest Gaia data on the outskirts of MWGCs, their tidal tails, and associated streams. We thoroughly review escape mechanisms from GCs and examine their relative contributions to the escape rate, ejection velocities, and escaper demographics. We show for the first time that three-body binary formation may dominate high-speed ejection from typical MWGCs, potentially explaining some of the hypervelocity stars in the MW. Due to their mass, black holes strongly catalyze this process, and their loss at the onset of observable core collapse, characterized by a steep central brightness profile, dramatically curtails three-body binary formation, despite the increased post-collapse density. We also demonstrate that even when born from a thermal eccentricity distribution, escaping binaries have significantly nonthermal eccentricities consistent with the roughly uniform distribution observed in the Galactic field.
\end{abstract}


\section{Introduction} \label{S:intro}
As some of the largest, densest, and oldest stellar systems in the Milky Way (MW), globular clusters (GCs) have long attracted interest as windows not only into stellar dynamics, but also as tracers of Galactic evolution. Of the ${\sim}170$ known MWGCs \citep{VasilievBaumgardt2021}, most in the Galactic disk likely formed within giant molecular clouds early in the MW's history \cite[e.g.,][]{PeeblesDicke1968}, but a substantial fraction of those in the Galactic halo likely originated in satellites that have since merged with the MW. Suggestively, halo GCs feature higher velocity dispersion but lower circular velocity than disk GCs, as well as lower metallicity and more retrograde orbits \citep[e.g.,][]{SearleZinn1978,RodgersPaltoglou1984,Zinn1985,Zinn1993,VanDenBergh1993,DaCostaArmandroff1995,Dinescu1999,Cote2000,ForbesBridges2010}. The MW's ongoing accretion of its satellites \citep[][]{Ibata1994,Martin2004,Belokurov2006} and their GCs \citep[e.g.,][]{Bellazzini2003,Forbes2004} reinforces this scenario.

Recent astronomical surveys, especially the Gaia survey \citep{GaiaEDR3}, have further revealed fine substructure in the MW halo, including numerous stellar streams \cite[for a recent review and catalog, see][]{Helmi2020,Mateu2023}.
These drawn-out associations of stars on similar orbits are likely debris from disrupted dwarf galaxies and their GCs, shorn off by Galactic tides during accretion by the MW. Gaia's exquisite kinematic data has firmly tied the origins of ${\sim}10$ especially thin streams to specific MWGCs \citep[e.g.,][]{Myeong2018b,Yuan2020,Bonaca2021,Ibata2021}. Twin strands of stars termed tidal tails also emanate from some MWGCs, most famously extending far from Palomar~5 \citep[e.g.,][]{Odenkirchen2001} into a full-fledged stream---but see \cite{Piatti2020} for a recent meta-analysis of many additional discoveries. These tails arise from the Coriolis effect in the rotating frame of a GC's orbit in the MW, which causes escapers headed toward (away from) the Galactic Center to speed ahead (trail behind) the GC \citep[e.g.,][]{BinneyTremaine2008}. Since tidal tails and stellar streams are excellent tracers of the MW's potential and recent merger history, their formation from stars escaping GCs is an essential topic in Galactic archeology.

While observational work on tidal tails and stellar streams is booming, theoretical study of stellar escape from GCs has a longer history. Numerous mechanisms (Section~\ref{S:escape_mechanisms}) cause high-speed escape, but low-speed escape via tides and two-body relaxation \cite[e.g.,][]{Ambartsumian1938,Spitzer1940,Chandrasekhar1943} generally dominates in realistic GCs subject to external tides. Yet escape in this case is complex. In principle, escape occurs once a star crosses the GC's tidal boundary, but this is neither guaranteed nor irreversible even for stars with high enough energies to do so. Back-scattering by encounters with other stars, the existence of periodic stellar orbits outside the tidal boundary, and the nonexistence of energy-based escape criteria for noncircular GC orbits add to the complexity (see the \hyperref[S:escape_conditions]{Appendix}). These nuances challenge detailed study of escaper properties, especially comparison between theory and observation. Yet the Gaia survey and its impact on Galactic archeology provide ample opportunity and motivation for renewed effort. Such comparison is necessary to validate and improve the escape physics in GC modeling and explore in-cluster origins of field stars. It may also help constrain otherwise difficult-to-measure GC properties, such as the extent of central black hole (BH) populations, supernova (SN) kick strengths, and the stellar initial mass function, which all affect ejection speeds and the GC evaporation rate \citep[e.g.,][]{Chatterjee2017,Weatherford2021}.

In this study, the first in a series, we lay essential groundwork for direct comparison between Gaia observations and GC models simulated with the \texttt{Cluster Monte Carlo} code (\texttt{CMC}). We do so by exploring in-depth the various mechanisms of escape from GCs, especially their role in \texttt{CMC}. We analyze escapers from our recent catalog of \texttt{CMC} models \citep{CMCCatalog}, focusing exclusively on properties at the time of removal from \texttt{CMC}, as opposed to the continued evolution of escaper trajectories in a full Galactic potential or any comparison to Gaia data---the focus of the second and third papers in this series, respectively.

We pay special attention to differences in escape before and after cluster \textit{core collapse}. Our usage of this term refers to the \textit{observable} transition from a flat (non-core-collapsed; NCC'd) to a steep (core-collapsed; CC'd) central surface brightness. This occurs upon dynamical ejection of the GC's central BH population, and corresponding transition from binary BH burning to binary white dwarf (WD) burning \citep[e.g.,][]{Chatterjee2013a,Kremer2019a,CMCCatalog,Kremer2021,Rui2021a}. \textit{Binary burning} refers to hardening of binaries in encounters with passing stars \citep[e.g.,][]{Heggie1975,Hills1975}; the potential energy released by hardening heats the stars and the binaries' centers of mass, halting collapse. BH binaries, being the most massive, have more potential energy so are stronger heat sources. Yet especially dense GCs quickly eject BHs via strong encounters. Central heating then relies on (less massive) WDs, weakening binary burning and allowing the core to observably collapse \citep[e.g.,][]{Kremer2019a,Kremer2021}. Note our usage is a refinement in detail to the conception of core collapse induced by the gravothermal instability \citep[e.g.,][]{Spitzer1987,HeggieHut2003} and halted by three-body binary formation (3BBF).  It also differs from the transient collapses characterizing \textit{gravothermal oscillations} \citep[e.g.,][]{HeggieHut2003}, in which the central regions of the cluster---especially BH populations \citep[e.g.,][]{Morscher2013,Morscher2015}---frequently and briefly contract via the gravothermal instability before re-expanding due to 3BBF. These mathematically chaotic oscillations occur throughout the GC's life, both before and after observable core collapse.

This paper is organized as follows. We first review numerous mechanisms of escape from GCs in Section~\ref{S:escape_mechanisms} before describing our \texttt{CMC} models in Section~\ref{S:models}. In Section~\ref{S:escaper_properties}, we examine escaper properties at the time of removal from \texttt{CMC}, emphasizing differences between CC'd and NCC'd GCs, as defined above. In particular, we analyze relative contributions from various escape mechanisms, distributions in ejection position and velocity, the escape rate over time, and properties of escaping binaries. We discuss the impact of 3BBF and limitations to our analysis in Section~\ref{S:disc} and summarize our findings and the future direction of this series in Section~\ref{S:summary}. In the \hyperref[S:escape_conditions]{Appendix}, we also include  a short review of static tides and a comparative discussion of several escape criteria commonly used in modeling.

\section{Escape Mechanisms} \label{S:escape_mechanisms}
Here we broadly review the numerous mechanisms contributing to escape from star clusters. Though \texttt{CMC} features many of these mechanisms (Section~\ref{S:escape_mechanisms_in_CMC}), our intent is not merely to describe those we do study, but also to provide a fuller picture of the escape landscape. Prior studies on cluster escape rarely discuss more than a few mechanisms at once, and a comprehensive review has yet to appear in the literature; both factors motivate thorough discussion.

Escape from GCs can be split into two categories: ejection and evaporation \citep[see][though our definitions are more expansive]{BinneyTremaine2008}. \textit{Ejection} occurs when a single dramatic event substantially increases the velocity of a cluster member all at once, potentially ejecting it with energy up to several times that necessary for escape. Events of this type include strong gravitational encounters---even collisions with other bodies---and assorted recoil kicks from stellar evolution, supernovae (SNe), tidal disruption events (TDEs), and gravitational-wave (GW)-driven mergers. In contrast, we define \textit{evaporation} mechanisms to be those \textit{incapable} of accelerating stars to speeds greatly in excess of the local escape speed. Mechanisms of this type include two-body relaxation and tidal stripping, which both operate on a macroscopic level---i.e., the key physics involves the cluster bulk rather than a single individual event or interaction. So they typically operate on timescales much greater than the dynamical timescale governing most (sudden) ejection mechanisms. Though evaporation mechanisms are less diverse than ejection mechanisms, they dominate the overall escape rate for realistic (tidally truncated) GCs, so we discuss them first.

\subsection{Evaporation} \label{S:evaporation_mechanisms}
The gradual evaporation of star clusters is a complex process given much attention over the years, including thorough discussions in classic field textbooks \citep[e.g.,][]{Spitzer1987,HeggieHut2003,BinneyTremaine2008}. We summarize key relevant features and note that our discussion differs slightly from earlier approaches; whereas many texts use \textit{evaporation} to refer to escape via two-body relaxation (or \textit{diffusion}), we expand this definition to include escape via global mass loss and time dependent tides.

\subsubsection{Two-body Relaxation} \label{S:relaxation}
In a gravitational $N$-body system, the motions of each body introduce granular, time dependent perturbations to the dominant, otherwise smooth underlying cluster potential. Via these fluctuations in the potential, the bodies exchange energy and momentum, causing each body to diffuse gradually and randomly through phase space. Though the exact trajectories obey the full set of coupled equations of motion, their numerical integration---as undertaken by direct $N$-body codes---is not especially instructive on a macroscopic level. The concept of two-body relaxation, specifically the Chandrasekhar theory of relaxation \citep{Chandrasekhar1942,Chandrasekhar1960,Spitzer1987,HeggieHut2003,Aarseth2008,BinneyTremaine2008} applies the simplifying approximation that the net diffusive effect of these perturbations is capturable as the sum of many \textit{weak}, \textit{impulsive}, and \textit{uncorrelated} two-body encounters. The meaning of \textit{weak} in this context is that each such encounter is distant enough to change the bodies' velocities $v$ by only a small amount ($\Delta v/v \ll 1$). Many such weak encounters cumulatively change $v$ by of order itself ($\Delta v/v \sim 1$) on the relaxation timescale---e.g., Equation~(2.62) in \citet{Spitzer1987}:
\begin{equation} \label{Eq:relaxation_timescale}
t_r \approx  \frac{0.065 \langle v\rangle^3}{G^2 \langle n\rangle \langle m\rangle^2 \ln\Lambda}.
\end{equation}
Here $\langle v\rangle$, $\langle n\rangle$, and $\langle m\rangle$ are the average velocity, number density, and stellar mass, respectively, while the Coulomb logarithm $\ln\Lambda\approx \ln (N/100)$ accounts for the range of impact parameters and depends on the initial mass function \citep[e.g.,][]{Freitag2006b,Rodriguez2018c}. Due to the spread in $v$, $n$, and $m$, the local $t_r$ is often several orders of magnitude longer in the GC's sparse halo than at its center, but the above is a reasonable fiducial timescale for the entire cluster.

As a random-walk process, relaxation drives the velocity distribution function (DF) toward thermal equilibrium, i.e., a Maxwellian DF. Bodies inevitably wander into the DF's high-speed tail, where a final weak encounter may push them beyond the cluster's local escape speed. Though later encounters will drive some back down to lower energy, rebinding them to the GC before long \citep[e.g.,][]{King1959}, many will eventually escape altogether. Crucially, the escape of these bodies evacuates the DF's high-speed tail, shifting the GC back away from equilibrium and encouraging more bodies to refill the tail. These, too, escape and the process repeats. The decline in cluster mass, aided by stellar evolution mass loss \citep[e.g.,][]{ChernoffWeinberg1990}, further enhances this cycle by continuously raising the GC's potential and decreasing the escape speed. So gravitational $N$-body systems never reach equilibrium\footnote{More generally, the gravothermal catastrophe \citep[e.g.,][]{LyndenBellWood1968,HeggieHut2003,BinneyTremaine2008} prevents \textit{any} finite gravitational system from reaching thermal equilibrium; cooling bodies simply sink deeper into the potential, causing them to speed up faster than they cool. This also prevents equipartition of (kinetic) energy, since heavier bodies sink via dynamical friction, causing them to heat rather than cool.} and have finite lifetimes dependent on $t_r$ \citep[e.g.,][]{Ambartsumian1938,Spitzer1940}. Back-scattering of \textit{potential escapers} back down to lower energy grows more efficient with lower $N$, increasing cluster lifetime \citep{Baumgardt2001}.

\subsubsection{Cluster Mass Loss} \label{S:cluster_mass_loss}
As noted above, the very act of losing mass raises (makes less negative) the cluster potential, thereby reducing the escape speed. So by unbinding stars already near the escape speed, mass loss can in principle be considered its own evaporation mechanism---especially when due to stellar evolution \citep[e.g.,][]{ChernoffWeinberg1990}, since it relies on no \textit{other} escape mechanisms to reduce the cluster mass \citep[see also Section 7.5.1 of][]{BinneyTremaine2008}. However, since cluster mass loss and two-body relaxation are both continuous, global phenomena, not discrete events like ejection during close encounters, it is impossible to definitively distinguish between these mechanisms when attributing the cause of escape for any specific body. Note the same degeneracy applies to time-dependent tides below, but unlike cluster mass loss and relaxation, these are often left out of cluster models.

\subsubsection{Sharply Time-dependent Tides} \label{S:tidal_stripping}
Star clusters also evaporate via global tidal effects---large-scale external perturbations to the cluster's gravitational potential. The simplest scenario, often used in modeling, is a cluster in a circular galactocentric orbit in an unchanging, spherical galactic potential. In this case, the galactic potential is static in the frame corotating with the orbit and imposes on the cluster a nonspherical tidal boundary (\textit{not} static since the cluster still loses mass). This eases escape via other mechanisms by lowering the escape energy (see Appendix~\ref{S:theoretical_energy_criterion}). However, without simultaneous internal energy exchange or mass loss via gravitational scattering or stellar evolution, there is no way for a static external tide to independently unbind cluster members. So to count as a truly distinct evaporation mechanism tides must be \textit{time dependent}.

Realistic tidal fields are time dependent in several ways. For example, the orbits of most MWGCs are both somewhat eccentric and inclined relative to the Galactic disk \citep[e.g.,][]{Baumgardt2019}. The former causes the Galactic tide to strengthen at perigalacticon, while the latter causes it to strengthen during passage through the Galactic disk \citep[e.g.,][]{Ostriker1972,SpitzerChevalier1973}. Similar perturbations occur during passage near any other mass within the Galaxy, including other GCs or giant molecular clouds \citep[e.g.,][]{Gieles2006}. Except in cases of nearly adiabatic time dependence, such as slow evolution of the galactic potential itself, any of these external perturbations to the cluster potential can induce \textit{tidal shocks} that heat the cluster through differential acceleration of individual stars relative to the cluster center \citep[see discussions in][]{Spitzer1987,HeggieHut2003,BinneyTremaine2008}.

Like two-body relaxation, tidal shocking can cause escape. Indeed, direct $N$-body models show that mass loss is significantly faster in GCs with highly eccentric and/or inclined orbits, in part due to shocking \citep[e.g.,][]{BaumgardtMakino2003,Webb2013,Webb2014,Madrid2014}. To significantly affect a star's orbit within the cluster, the shocks must be relatively impulsive (occur on a timescale less than the star's orbital period). This is especially likely in the cluster halo, where the crossing time is longer. Shock heating may even exceed heating from two-body relaxation in the halos of disk-crossing GCs \citep[e.g.,][]{KundicOstriker1995,GnedinOstriker1997,Gnedin1999a,Gnedin1999b}, but the latter dominates in all but the most massive  GCs \citep[e.g.,][]{FallZhang2001,McLaughlinFall2008,PrietoGnedin2008}.

\subsection{Ejection} \label{S:ejection_mechanisms}
There are many ways to eject objects from star clusters in individual, dramatic, potentially violent events. So it serves us well to further divide ejection mechanisms into three classes: strong encounters, (near-)contact recoil, and stellar evolution recoil. Fundamentally, the second is simply a more extreme extension of the first that may involve elements of the third. In general, keep in mind these mechanisms are not mutually exclusive, though stellar evolution recoil is conceptually distinct by not relying on gravitational scattering.

\subsubsection{Strong Encounters} \label{S:strong_encounters}
Strong gravitational scattering interactions are some of the most well-studied ejection mechanisms, featuring close passages between two or more bodies or bound hierarchies (e.g., binaries and triples) that may apply large dynamical kicks to one or more bodies. Kick magnitudes have large variance, depending strongly on the relative speed and orientations of the interacting bodies/hierarchies. Strong encounter ejection mechanisms include the following:
\begin{itemize}
\item \textit{Strong two-body encounters}: Weak perturbations dominate the \textit{average} rate of energy change experienced by cluster members in two-body scattering, enabling the classic relaxation theory described above. The result is a relatively smooth random walk in each body's orbital energy, unable to induce velocities much larger than the cluster's local escape speed. However, the true random walks are sharper and more granular---more conducive to high-speed ejection. In the right circumstances (e.g., a flyby of a much more massive object on a similar trajectory), a single close encounter can be much stronger---up to $\Delta v/v \approx 3$ \citep[e.g.,][]{Henon1969}. Though rare, such encounters would have outsize influence on the high-velocity end of the ejection speed distribution, and are not captured by standard two-body relaxation. The impact of strong two-body encounters is greater in isolated clusters, where relaxation is less effective at causing escape \citep{Henon1960,Henon1969,SpitzerShapiro1972}.
\item \textit{Three-body encounters between singles}:
Strong encounters between \textit{three} separate bodies are rarer than between two. Yet they dominate new binary formation \citep[e.g.,][]{HeggieHut2003} since two of the bodies often bind together, with increasing probability for stronger encounters \citep{AarsethHeggie1976}. The potential energy released in binding accelerates the leftover single and the new binary's center of mass. These kicks can easily eject the single from the cluster. (Binary ejection is rare since those formed this way are biased to much higher mass than the leftover single and so receive smaller kicks under momentum conservation; see Section~\ref{S:3BBF}.) Even the encounters that do \textit{not} form binaries can still cause ejection, albeit at lower speeds due to the lesser release of potential energy. Yet strong encounters between three singles have attracted little attention due to their rarity for typical stellar masses in even dense GCs. This reasoning neglects the encounter rate's exceptional sensitivity to the masses and velocities of the species involved. Central BH populations in GCs especially enhance three-body binary formation \citep[e.g.,][]{Kulkarni1993,OLeary2006,Morscher2013,Morscher2015}. In Section~\ref{S:ejecta_by_escape_mechanisms}, we show this enhancement may allow 3BBF to dominate high-speed stellar escape from MWGCs at present.

\item \textit{Binary--single encounters}: Three-body encounters between a binary and a single also commonly lead to large dynamical kicks. In particular, the flyby of a single star near a sufficiently hard (compact) binary tends to further harden the binary, the released potential energy again boosting the speeds of the single and binary center of mass \citep[e.g.,][]{Heggie1975,Hills1975,SigurdssonPhinney1993}. When a three-body interaction involving a hard binary features a high mass ratio, ejection speeds of several hundred kilometers per second are possible \citep[e.g.,][]{Gvaramadze2009} and help explain the high observed velocities of O- and B-type stars in the MW \citep[e.g.,][]{Fujii2011}.
\item \textit{Binary--binary encounters}: Strong four-body encounters involving two binaries are rarer than binary--single encounters due to the limited binary fraction in a GC's dense core \citep[e.g.,][]{Milone2012}. However, by providing additional binding energy to transfer into orbital speeds and another star to prolong resonance interactions featuring especially close pericenter passages, binary--binary scattering can cause very high-speed ejections of order several hundred kilometers per second \citep[e.g.,][]{LeonardDuncan1988,LeonardDuncan1990,Gualandris2004}, potentially above $10^3{\rm\ km\ s}^{-1}$ \citep[e.g.,][]{Leonard1991}.
\item \textit{Higher-$N$ encounters}: The above reasoning extends to paired interactions between larger-$N$ bound hierarchies (triple--single, triple--binary, etc.), higher-multiplicity single interactions (e.g., four-body binary formation), or some combination (e.g., binary--single--single)---in principle up to the size of the GC. But as hierarchy size and interaction multiplicity grow, the encounter rate rapidly diminishes in typical GCs, too dense to accommodate durable hierarchies yet too diffuse to feature appreciable rates of high-multiplicity strong encounters \citep[e.g.,][]{Atallah2023}. From a practical standpoint, direct $N$-body codes implicitly incorporate these physics by fully integrating trajectories, while introducing higher-$N$ encounters into alternatives like Monte Carlo or Fokker--Planck codes would quickly erode their main advantage---computational speed. So while higher-$N$ encounters may allow slightly more or higher-speed ejections---due to the remote chance of chained gravitational slingshots like those used in spacecraft maneuvers---any increase is likely small.
\item \textit{Unstable triple disintegration}: Triples and larger hierarchies are ephemeral in typical GCs but present a unique ejection mechanism beyond encounters with other bodies. Triples can be unstable to gravitational perturbations or stellar mass loss \citep[e.g.,][]{MardlingAarseth2001}. When instability causes them to disintegrate (typically into a binary and single), the released binding energy can accelerate the separating bodies by tens to hundreds of kilometers per second \citep{Toonen2022}.
\end{itemize}

\subsubsection{(Near-)Contact Recoil} \label{S:near_contact_recoil}
Ejection physics is more complex when objects pass \textit{so} close that tides, internal stellar processes, and/or relativistic effects are relevant. Such strong encounters often feature direct contact between two bodies or nearly so. As with the strong encounters above, the encounter rate increases with number density and stellar mass, but the latter's influence is amplified since the intrinsic size of each body, which typically increases with mass, now matters. Binaries also greatly enhance the rates due to their large cross sections and since strong encounters involving them are often \textit{resonant}, featuring chaotic series of many separate pericenter passages \citep[e.g.,][]{Bacon1996,Fregeau2004}.
\begin{itemize}
\item \textit{Direct physical collisions}: From the standpoint of distance, the strongest gravitational interaction two bodies may experience is a rare head-on collision. Though much of the released gravitational energy goes into the internal energy of the collision remnant(s), asymmetric mass ejection during the collision can kick the remnant(s) and, by momentum conservation, any other bodies in the interaction. Hydrodynamic simulations suggest such kicks can reach ${\sim} 10{\rm\ km\ s}^{-1}$ in star--star collisions \citep{Gaburov2010} or even up to ${\sim} 100{\rm\ km\ s}^{-1}$ in BH--star collisions \citep{Kremer2022_BH}. In general, however, the kick speeds depend sensitively on the exact species and processes involved.

\item \textit{TDEs}: Less extreme than a physical collision is a TDE, in which a star passes close enough to another body for that body to strip away some of the star's mass, potentially even destroying it entirely \citep[e.g.,][]{Rees_1988,Kremer2019_TDE,SamsingVenumadhav2019,FragionePerna2021}. If the mass loss is asymmetric, the TDE also kicks the remnant(s). Hydrodynamic simulations suggest mass loss during the tidal capture of a star into a binary with a BH (or complete disruption of the star) can kick the new binary's center of mass (or, for full disruptions, the leftover lone BH) by ${\sim} 10$--$100{\rm\ km\ s}^{-1}$ \citep{Kremer2022_BH}. The kick is typically highest for more-penetrating encounters, which cause more mass loss. Similar kicks can apply directly to the unbound stellar remnant itself and more extreme kicks at several hundred kilometers per second can occur in TDEs during encounters between tight stellar binaries and BHs \citep{Ryu2023} or even at ${\gtrsim} 10^3{\rm\ km\ s}^{-1}$ in TDEs involving $10^2{-}10^4~M_{\odot}$ intermediate-mass BHs \citep[IMBHs;][]{Kiroglu2022b}. Kicks from asymmetric mass loss have also been studied in the context of planets \citep{Faber2005, Guillochon2011, Liu2013}, main-sequence (MS) stars \citep{Manukian2013,Gafton2015,Ryu2020a,Ryu2020b,Ryu2020c}, WDs \citep{Cheng2013}, and neutron stars \citep[NSs;][]{Rosswog2000,Kyutoku2013,Kremer2022_NS}. 

\item \textit{GW-driven mergers}: Finally, mergers driven by GW dissipation, in particular BH mergers, emit GWs asymmetrically, often kicking the remnant by tens to hundreds of kilometers per second \citep[e.g.,][]{Bonnor1961,Peres1962,Bekenstein1973,Favata2004,Merritt2004a,LoustoZlochower2008,LoustoZlochower2009,LoustoZlochower2011,Lousto2010,Lousto2012}. This notably hinders IMBH growth via BH mergers in star clusters, as the kick often ejects the remnant except at extreme mass ratios---such as a low-mass BH merging with a preexisting IMBH \citep[e.g.,][]{Bockelmann_2008,Moody_2009,Morawski_2018,Rasskazov_2020,FragioneLoeb2021,Sedda_2021,FragioneLoeb2022,FragioneKocsis2022,Gonzalez_2022,Maliszewski_2022}.
\end{itemize}

\subsubsection{Interactions with IMBHs} \label{S:IMBHs}
IMBHs in the range of $10^2$--$10^4\,M_\odot$, if present in a GC, may strongly affect cluster evolution and ejection speeds \citep[e.g.,][]{BaumgardtMakino2005,Baumgardt2017}. Interactions with IMBHs do not necessarily constitute a separate ejection mechanism, since they are just a more extreme case of strong encounters or TDEs involving typical stellar BHs. Yet this channel is worth highlighting for its potential to help identify IMBHs in GCs.

The process of disrupting a stellar binary via a binary--single strong encounter with a massive BH (classically a supermassive BH) is often known as the Hills mechanism \citep{Hills1988}, though note this is merely an extreme case of an ordinary stellar binary--single encounter. The tidal radius of the BH is $r_t \sim (M/m)^{1/3} a$, where $M$ is the BH mass, $m=m_1+m_2$ the binary mass, and $a$ the binary semi-major axis. When the binary comes within distance $r_t$ of the BH, it is typically disrupted, leaving one star bound to the BH and ejecting the other at high speed. Recently, \cite{FragioneGualandris2019} demonstrated that for typical stellar masses and binary semi-major axes in the core of a GC, a $10^3\,M_\odot$ IMBH at the GC's center would eject stars at speeds most commonly near $700{\rm\ km\ s}^{-1}$ and up to ${>}10^3{\rm\ km\ s}^{-1}$. An earlier rate analysis by \cite{Pfahl2005} and reexamined by \cite{FragioneGualandris2019} suggests that a typical GC hosting a $10^3\,M_\odot$ IMBH would eject such stars at a rate of ${\approx}0.1{\rm\ Myr}^{-1}$, yielding a highly significant contribution to the high-speed ejection distribution of a GC. As discussed above, TDEs by IMBHs can also lead to high-speed stellar ejections \citep{Kiroglu2022b}. Possible instances of each of these mechanisms have already been observed \citep[e.g.,][]{Gualandris2007,Lin2018}. So, while we do not examine this channel in this study, high-speed ejecta from MWGCs may eventually provide a valuable constraint on retention of IMBHs in specific MWGCs.

\subsubsection{Stellar Evolution Recoil} \label{S:stellar_evolution_recoil}
Stellar evolution preceding/during the birth of compact objects can also cause ejection. In particular, SNe that form NSs and BHs may feature asymmetric ejection of matter, giving their remnants impulsive recoil kicks known as SNe kicks. By momentum conservation, these kicks may also eject binary companions to SNe progenitors---in fact, \cite{Blaauw1961} originally proposed SNe kicks to explain speedy O- and B-type stars, suggesting they were once binary companions to SNe progenitors. Asymmetric stellar winds may even propel WD progenitors. We detail several subtypes of stellar evolution recoil below.

\textit{SNe kicks in isolation}: The simplest case of SN-induced ejection from clusters occurs when the progenitor star exists in isolation (i.e., not a member of a binary or higher-order bound hierarchy). In such cases, the SN kick simply changes the velocity of the compact remnant itself, often by enough to eject it from the cluster, leaving it as a lone dark object.
\begin{itemize}
    \item \textit{NS SNe kicks} are the strongest and most observationally supported form of stellar evolution recoil. Empirical evidence for these kicks exists in the high velocity dispersion of the pulsars \citep[e.g.,][]{Lyne1982,LyneLorimer1994,Hansen1997,Arzoumanian2002,Hobbs2005} or otherwise-observable NSs of the MW \citep[e.g., those detectable via nebular bow shocks;][]{Cordes1993}. Typical kick magnitudes are hundreds of kilometers per second, which can be strong enough to eject NSs from galaxies, let alone star clusters. Consequently, most NSs escape their host clusters at birth.
    \item \textit{BH SN kick} magnitudes are more uncertain but are plausibly lower than NS kicks due to additional fallback of ejected mass onto the SN remnant. Indeed, the positions and velocities of BH X-ray binaries (XRBs) in the MW suggest that while at least some BHs receive SNe kicks of ${\sim} 100{\rm\ km\ s}^{-1}$, many likely receive lesser or negligible kicks \citep[e.g.,][and references therein]{Repetto2017}. While such speeds are sufficient to eject many BHs, the observed range of XRB velocities is consistent with significant BH retention in GCs \citep[e.g.,][]{Chatterjee2017}. Further observational support for such retention has been found in GCs' surface brightness and velocity dispersion profiles and internal mass segregation \citep[e.g.,][]{Merritt2004b,Mackey2007,Mackey2008,Peuten2016,ArcaSedda2018,Askar2018,Kremer2018,Kremer2019a,CMCCatalog,Weatherford2018,Weatherford2020,Zocchi2019,Rui2021a}. BH microlensing events may soon provide more detailed constraints on natal kick speeds \citep[e.g.,][]{Andrews2022}.
\end{itemize}

\textit{SNe kicks in binaries}: Various escape outcomes are possible when an SN takes place in a binary \citep[see, e.g., Appendix A1 of][for a quantitative description]{Hurley2002}. Depending on its strength and direction, the raw kick experienced by the SN remnant may be weak enough to drag its binary companion with it, resulting in a weaker effective kick to the center of mass of the binary, still bound together but with altered orbital parameters. The raw SN kick may even be strong enough to fully unbind the binary, reducing the SN remnant's final speed but applying an \textit{induced kick} to the companion. This leads to the following ejection scenarios:
\begin{itemize}
    \item \textit{Binary ejection}: In this case, the raw SN kick is not enough to unbind the binary but is large enough to eject it from the cluster, for example, as an XRB or even doubly compact binary. 
    \item \textit{Binary disruption}: Alternatively, kicks great enough to unbind the binary may also eject either the SN remnant or its companion, often both. A key difference between this scenario and the isolated SN ejection scenario above is that while the SN remnant again escapes on its own, it does so at reduced speed, as unbinding the binary consumes some of the kick's energy.
    \item \textit{SN-induced (near-)contact recoil}: Finally, when the SN kicks the remnant toward its companion, it may physically collide with or tidally disrupt the companion, yielding similar scenarios to the dynamically induced mechanisms discussed in Section~\ref{S:near_contact_recoil}.
\end{itemize}

\textit{WD kicks}: WDs may also experience stellar evolution recoil under certain circumstances. In particular, asymmetric mass loss during the asymptotic giant branch (AGB) phase of stellar evolution may gently propel a WD progenitor. (Note this may also affect a BH or NS progenitor during its AGB phase, but less impactfully since the SN kick shortly thereafter would likely dwarf such propulsion.) If true, this would help explain a variety of WD observations, including an underabundance in open clusters \citep[e.g.,][]{Weidemann1977,Kalirai2001,Fellhauer2003}, wide spatial distribution, and low velocity dispersion in GCs \citep[e.g.,][]{Heyl2007,Heyl2008a,Heyl2008b,Calamida2008,Davis2008}, and unexpectedly wide semi-major axes among WD-containing binaries observed with Gaia \citep{ElBadryRix2018}. Each finding tenuously supports nonzero WD birth kicks, though no more than a few kilometers per second. This would only be large enough to eject WDs already relatively near their host GC's local escape speed. Alternatively, some actively accreting WDs may experience \textit{failed} Type 1a SNe capable of larger kicks of ${\sim}100{\rm\ km\ s}^{-1}$ \citep[e.g.,][]{Jordan2012}, though this kick magnitude is highly uncertain \citep[e.g.,][]{Kromer2013}.

\section{Cluster Models} \label{S:models}
We simulate MWGCs using \texttt{CMC} (for \texttt{Cluster Monte Carlo}), a H\'{e}non-type \citep{Henon1971a,Henon1971b} Monte Carlo code for star cluster modeling \citep[see][for the most recent and thorough overview]{CMCRelease}. \texttt{CMC} includes prescriptions for numerous physics essential to the evolution of massive GCs, including stellar evolution (comments below), two-body relaxation \citep{Joshi2000,Pattabiraman2013}, galactic tidal fields \citep[][]{Joshi2001,Chatterjee2010}, three-body binary formation \citep{Morscher2013,Morscher2015}, physical collisions \citep{Fregeau2007}, and strong three- and four-body scattering \citep{Fregeau2003,Fregeau2007} performed with the small-$N$ direct integrator \texttt{fewbody}, which includes post-Newtonian dynamics \citep{Fregeau2004,Antognini2014,AmaroSeoane2016,Rodriguez2018a,Rodriguez2018b}. \texttt{CMC} also allows for two-body binary formation through GW and tidal capture \citep{CMCCatalog,Ye2022}, but only the former is included in the models used here.

\floattable
\begin{deluxetable*}{r|lcc|cccccccccc}
\tabletypesize{\footnotesize}
\tablewidth{0pt}
\tablecaption{Initial/Final Cluster Properties and Final Population Counts \label{table:model_properties}}
\tablehead{
    \colhead{} &
    \colhead{Simulation from the} &
    \colhead{$r_v$} &
    \colhead{$R_g$} &
    \colhead{$r_c$} &
    \colhead{$r_h$} &
    \colhead{$r_t$} &
    \colhead{$v_{\rm esc,0}$} &
    \multicolumn{6}{c}{Final Population Counts} \\[-0.2cm]
	\colhead{} &
	\colhead{\texttt{CMC} Cluster Catalog} &
	\colhead{(pc)} &
	\colhead{(kpc)} &
	\colhead{(pc)} &
	\colhead{(pc)} &
	\colhead{(pc)} &
	\colhead{(${\rm km\ s}^{-1}$)} &
	\colhead{MS} &
	\colhead{G} &
	\colhead{WD} &
	\colhead{NS} &
	\colhead{BH} &
	\colhead{Total} 
}
\startdata
1  & \textsc{N8-RV0.5-RG2-Z0.1}  & 0.5 &  2 & 0.11 & 3.89 &  28.9 & 32.0 & 316,698 & 1612 & 62,257 & 248 &   1 & 365,516 \\
2  & \textsc{N8-RV0.5-RG8-Z0.1}  & 0.5 &  8 & 0.14 & 4.87 &  82.7 & 35.0 & 535,190 & 2073 & 78,615 & 278 &   0 & 591,369 \\
3  & \textsc{N8-RV0.5-RG20-Z0.1} & 0.5 & 20 & 0.18 & 5.28 & 156.6 & 35.4 & 598,223 & 2214 & 83,547 & 307 &   1 & 656,581 \\ \hline
4  & \textsc{N8-RV1-RG2-Z0.1}    & 1   &  2 & 1.03 & 3.97 &  31.3 & 24.9 & 437,164 & 2028 & 73,153 & 202 &  20 & 489,449 \\
5  & \textsc{N8-RV1-RG8-Z0.1}    & 1   &  8 & 1.22 & 4.85 &  85.7 & 26.2 & 617,548 & 2306 & 84,467 & 237 &  25 & 673,931 \\
6  & \textsc{N8-RV1-RG20-Z0.1}   & 1   & 20 & 1.44 & 5.26 & 160.5 & 25.6 & 659,786 & 2365 & 86,974 & 243 &  33 & 717,040 \\ \hline
7  & \textsc{N8-RV2-RG2-Z0.1}    & 2   &  2 & 2.62 & 5.86 &  30.4 & 16.8 & 400,030 & 1850 & 66,439 & 112 &  72 & 447,165 \\
8  & \textsc{N8-RV2-RG8-Z0.1}    & 2   &  8 & 2.79 & 7.00 &  86.9 & 20.3 & 654,814 & 2346 & 85,668 & 160 &  89 & 710,590 \\
9  & \textsc{N8-RV2-RG20-Z0.1}   & 2   & 20 & 2.93 & 7.58 & 161.7 & 20.6 & 730,566 & 3246 & 74,113 & 166 & 119 & 773,267 \\ \hline
10 & \textsc{N8-RV4-RG2-Z0.1}    & 4   &  2 & \multicolumn{10}{c}{\textit{Disrupted}} \\
11 & \textsc{N8-RV4-RG8-Z0.1}    & 4   &  8 & 4.78 & 11.1 &  87.4 & 16.8 & 661,605 & 2358 & 84,716 &  75 & 269 & 715,520 \\
12 & \textsc{N8-RV4-RG20-Z0.1}   & 4   & 20 & 4.99 & 11.6 & 163.2 & 16.7 & 693,645 & 2437 & 87,763 & 106 & 285 & 749,357 \\
\enddata
\tablecomments{\footnotesize Initial virial radius $r_v$, Galactocentric distance $R_g$, and final core, half-mass, and tidal radii ($r_c$, $r_h$, and $r_t$, respectively), central escape speed $v_{\rm esc,0}$, and in-cluster population counts for MS = main-sequence stars, G = giants, WD = white dwarfs, NS = neutron stars, BH = black holes, and their combined total. Simulation 10 is excluded from the final counts since it disrupted long before a Hubble time.}
\end{deluxetable*}

\vspace{-23pt}
We limit our analysis to a subset of 12 GC simulations (see Table~\ref{table:model_properties}) from our larger \texttt{CMC} Cluster Catalog \citep{CMCCatalog}. Each begin with an initial number of particles (single stars plus binaries) $N_i=8\times 10^5$ and cluster metallicity $Z/Z_\odot=0.1$. The chosen $N_i$ yields GCs near the average of the present-day MWGC mass distribution \citep[e.g.,][]{Mandushev1991,BaumgardtCatalog2020} and $Z$ is similarly typical of MWGCs \cite[e.g.,][2010 edition]{HarrisCatalog1996}. In these simulations, we only vary the initial virial radius $r_v/{\rm pc}\in [0.5,1,2,4]$ and Galactocentric distance $R_g/{\rm kpc}\in [2,8,20]$, also designed to capture the spread of MWGCs \citep{CMCCatalog}. Yet due to the highly detailed nature of this study focusing on escape mechanisms and demographics, we devote most of our attention to two archetypal GC models: the two most representative of typical CC'd and NCC'd MWGCs (models 2 and 8 in Table~\ref{table:model_properties}, respectively). These differ in only their initial virial radius $(r_v/{\rm pc}=0.5,2)$, which controls the initial density and hence the timescales for relaxation and dynamical ejection of central BHs.
Recall the latter induces core collapse as defined in Section~\ref{S:intro}---the transition from a flat to a steep central surface brightness and long-term contraction of the core radius $r_c$. In the following sections, we use the \textit{theoretical} density-weighted $r_c$ from \cite{CasertanoHut1985}. Not only does this shrink when the central surface brightness steepens, it is also more sensitive to transient density oscillations among central compact objects than measures based solely on surface brightness or cumulative luminosity profiles \citep{Chatterjee2017}.

In each simulation, the randomized initial positions and velocities derive from a \cite{King1966} profile with concentration $w_0 = 5$. Stellar masses (primary mass $m_p$, in the case of a binary) draw from the standard \cite{Kroupa2001} initial mass function from 0.08--$150\, M_\odot$. Binary sampling proceeds by randomly assigning secondaries to $N\times f_b$ stars, independent of radial position or mass, where $f_b=5\%$. Secondary masses adopt a uniform mass ratio $q\in [0.08/m_p,1]$ and binary orbital periods draw from a distribution flat in log-scale \citep[e.g.,][]{DuquennoyMayor1991}, where the orbital separations range from near contact to the hard/soft boundary. Binary eccentricities are thermal \citep[][]{Heggie1975} and we allow each GC simulation to evolve to a final time of $14$~Gyr.

Simulations from the \texttt{CMC} Cluster Catalog implement stellar evolution via the single/binary-star evolution codes \texttt{SSE}/\texttt{BSE} \citep{Hurley2000,Hurley2002}, with recent prescriptions for wind-driven mass loss, compact object formation, and pulsational-pair instability \citep[see][]{CMCCatalog}. While minor enhancements have since been and continue to be implemented into the latest publicly available version of \texttt{CMC}---now via the stellar evolution code \texttt{COSMIC} \citep{Breivik2020}, an \texttt{SSE/BSE} derivative---such enhancements are largely irrelevant for this study focusing on escape mechanisms. When ongoing stellar evolution after escape becomes relevant for constructing extra-tidal observables in our follow-up work (Weatherford et al. 2023b, in preparation), we continue evolving escapers with \texttt{COSMIC}.

\subsection{Tidal Truncation and Escape in \texttt{CMC}} \label{S:tidal_truncation}
By default, \texttt{CMC} models are tidally limited under the assumption that the clusters circularly orbit the Galactic Center with circular speed $V_G = \Omega R_G = 220\,{\rm km}\,{\rm s}^{-1}$, typical of MWGCs \citep[e.g.,][]{BinneyTremaine2008}. \texttt{CMC} defines its tidal radius as $r_t\equiv r_{\rm J}$, utilizing the more accurate calculation of the Jacobi radius $r_{\rm J}$ for a realistic logarithmic Galactic potential (see Appendix~\ref{S:theoretical_energy_criterion}), leading to \citep[e.g.,][]{Spitzer1987,BaumgardtMakino2003,Chatterjee2010}:
\begin{align} \label{Eq:rtCMC}
r_t(t) \equiv \left(\frac{M_C(t)}{2M_G} \right)^{1/3} R_G = \left( \frac{GM_C(t)}{2V_G^2} \right)^{1/3} R_G^{2/3}.
\end{align}
Note the explicit time dependence through the cluster mass, which decreases via both stellar escape and stellar winds.

To determine escape, \texttt{CMC} uses an energy-based criterion \citep{Giersz2008} tuned to reproduce well the escape rate from direct $N$-body models. This criterion modifies the standard energy criterion (Appendix~\ref{S:theoretical_energy_criterion}) to account for back-scattering of \textit{potential} escapers down to lower energies before they can cross the tidal boundary. Specifically, \texttt{CMC} immediately removes any particles with specific energy greater than the cluster potential at $r_t$ times an order unity factor,
\begin{equation} \label{Eq:Giersz_energy_criterion}
E > \alpha \phi_c\left(r_t\right),
\end{equation}
where the factor $\alpha$, tuned by comparison to direct $N$-body models, is given by
\begin{equation} \label{Eq:Giersz_alpha}
\alpha \equiv 1.5 - 3 \left( \frac{\ln\Lambda}{N} \right)^{1/4}.
\end{equation}
Here, $\ln\Lambda = \ln\left(\gamma N\right)$ is the Coulomb logarithm. In \texttt{CMC}, $\gamma$ defaults to 0.01 \citep[see also][]{Freitag2006b,Rodriguez2018c,CMCRelease}. Numerical testing \citep{Giersz2008,Chatterjee2010} has shown that Equation~(\ref{Eq:Giersz_energy_criterion}) produces significantly better agreement to direct $N$-body simulations than alternatives, such as simply stripping stars with apocenters beyond the tidal boundary (for further discussion, see Appendix~\ref{S:code_escape_criteria}).

\subsection{Escape Mechanisms in \texttt{CMC}} \label{S:escape_mechanisms_in_CMC}
\texttt{CMC} features many of the escape mechanisms in Section~\ref{S:escape_mechanisms}, as neatly listed in Table~\ref{table:escape_mechanisms_in_CMC}. The implementation of two-body relaxation and several types of strong interactions---binary--single and binary--binary encounters, 3BBF, and physical collisions---features extensively in \cite{CMCRelease} and references therein. Regarding 3BBF especially, we note that while earlier \texttt{CMC} models \citep[since][]{Morscher2013} only allowed 3BBF between three BHs, the \texttt{CMC} Cluster Catalog models allow this process between any three bodies, as is generally possible (though one or more BHs are usually involved; see Section~\ref{S:closer_look_at_3bb}). Meanwhile, physical collisions in \texttt{CMC} use a simplified sticky-sphere approximation, occurring whenever the radii of any two bodies (except NSs or BHs) overlap, at which point the bodies merge with partial mass loss if one is a WD and no mass loss otherwise. A collision between a BH/NS and any other star entirely destroys the latter without affecting the NS/BH mass. In all cases, the collisions conserve total momentum and assume any mass loss is spherically symmetric, so no kicks from asymmetric matter ejection apply to the collision remnant. This guarantees the remnant has speed $v<v_{\rm esc}$ unless one of the colliding bodies already had $v>v_{\rm esc}$, so we do not count collisions as a true escape mechanism in \texttt{CMC}. Related \texttt{CMC} prescriptions for TDEs are in development and not included either. Similarly, while \texttt{CMC} allows for time dependent tidal fields \citep[input at simulation start as a table describing a time-varying tidal tensor, e.g.,][]{Rodriguez2022a}, we leave examination of these effects for later studies.

As for other scattering-driven escape mechanisms, \texttt{CMC} does not yet account for strong two-body encounters or higher-$N$ hierarchies/interaction multiplicities (e.g., triples or four-body binary formation). While triples do form in \texttt{fewbody}, \texttt{CMC} simply breaks them apart at the end of each time step, so they do not participate in global dynamics \citep[for further discussion, see][]{Fragione2020}. Since \texttt{CMC} passes the released binding energy to neighboring stars this technically \textit{can} cause escape---implicitly lumped in with relaxation in this study. It is merely doubtful that such an artificial approach faithfully mimics ejection via triple disruption, especially the ejection speed distribution.

\texttt{CMC} does include prescriptions for GW-driven mergers and SNe kicks, however. Specifically, \texttt{CMC} treats binary BH mergers---including product mass, spin, and recoil kick---with formulae, fitted to numerical relativity simulations, from \cite{GerosaKesden2016}. For further details, see \cite{Rodriguez2018a,Rodriguez2018b}. SNe prescriptions draw from the \textit{rapid} model by \cite{Fryer2012} for standard iron core-collapse SNe (CCSNe) and from  \cite{Belczynski2016} for pair-instability and pulsational-pair-instability SNe. Electron-capture SNe (ECSNe) from several channels---including accretion- or merger-induced collapse of WDs---also feature in \texttt{CMC} \citep[][]{Ye2019}. The NS recoil kicks $V_{\rm NS}$ draw from Maxwellians with disperions $\sigma=265\,{\rm km\,s}^{-1}$ (CCSNe) or $\sigma=20\,{\rm km\,s}^{-1}$ (ECSNe). The BH kicks follow the NS kick distribution for CCSNe, but reduced according to the fraction $f_{\rm fb}$ of the progenitor's stellar envelope that falls back onto the BH remnant: $V_{\rm BH} = V_{\rm NS}(1-f_{\rm fb})$. For further details, see \cite{CMCCatalog} and note in contrast that since evidence for significant WD kicks remains tenuous, \texttt{CMC} does not include them. Finally, since stellar evolution in \texttt{CMC} occurs concurrently with \texttt{fewbody}, SNe kicks conserve momentum in the integrator, allowing for more complex ejection mechanisms like SN-induced binary ejection or disruption and SN-induced (near-)contact recoil.

\begin{figure*} 
\centering
\includegraphics[scale=0.64]{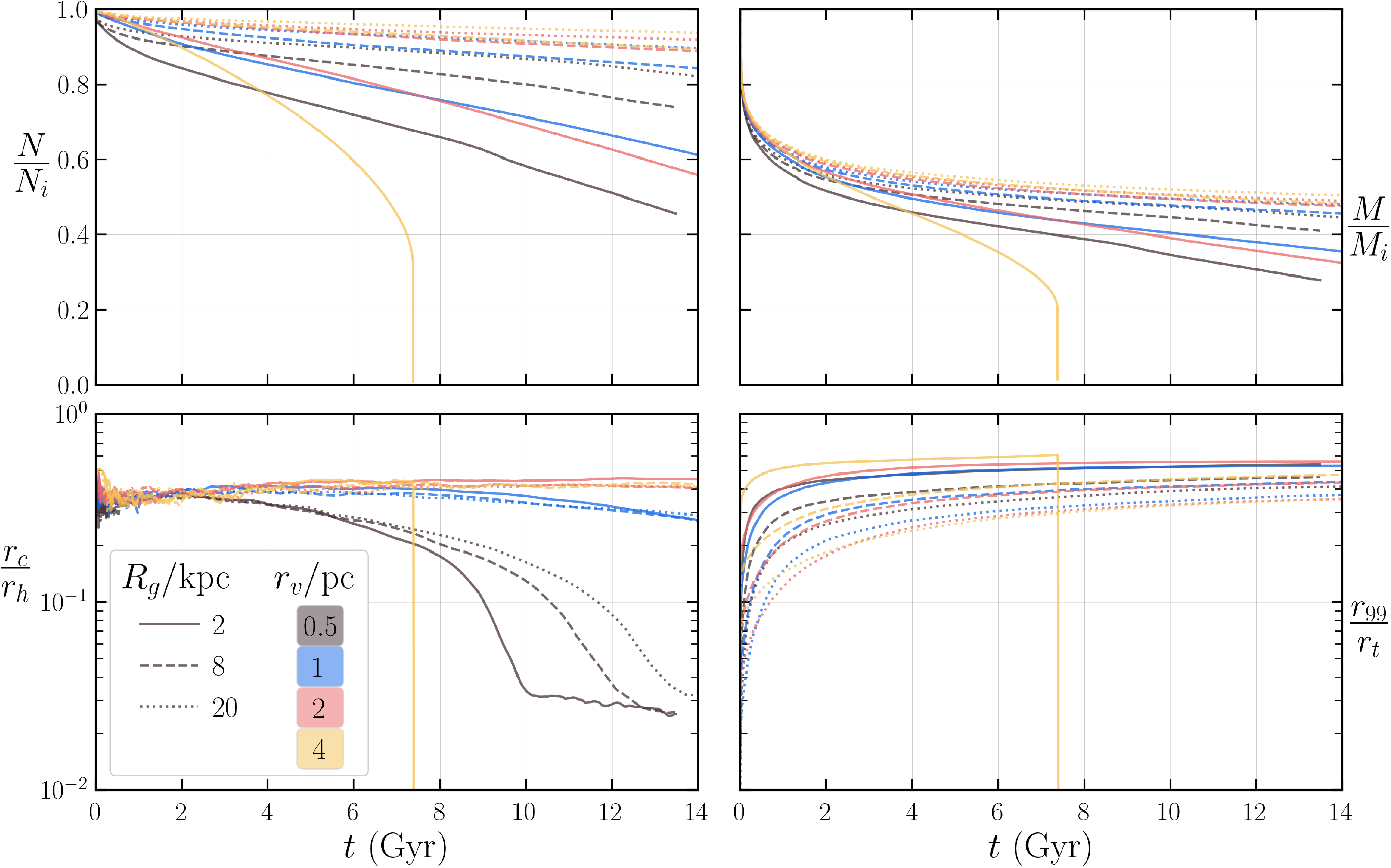}
\caption{Evolution of cluster models over time. Top left: number of particles relative to the initial number of particles. Top right: cluster mass relative to the initial cluster mass. Bottom left: core radius over half-mass radius (rolling average). Bottom right: 99\% Lagrange radius over tidal radius (rolling average). In all panels, line styles indicate the model's Galactocentric distance while line colors indicate the model's virial radius.}
\label{fig:Fig1}
\end{figure*}

\subsubsection{Classifying Escape Mechanisms in Post-processing} \label{S:classifying_escape_mechanisms}
In Section~\ref{S:escaper_properties}, we classify escapers by escape mechanism based on detailed \texttt{CMC} output---notably logs of all \texttt{fewbody} interactions, 3BBF, collisions, BH-forming SNe, BH mergers, and SN-induced binary disruption events. To determine the escape mechanism for each escaper (single or binary), we search each of the aforementioned log files for an event involving that body (component bodies in the case of a binary) at the time of removal from \texttt{CMC}. If the body or binary components do not appear in any of the log files at the appropriate time, process of elimination attributes the cause of escape to relaxation---implicitly coupled with static tides and triple disintegration, as explained earlier. Since simulations from the \texttt{CMC} Cluster Catalog do not have logs of NS-forming SNe, which are being added for future models, we lump all escaping NSs together without specifying a mechanism. Note, however, that SNe \textit{do} eject the vast majority of NSs in our models.

Also note that the ejection mechanisms are not mutually exclusive. For example, SNe and BH mergers may both take place, sometimes several times, during a single \texttt{fewbody} interaction, each applying separate kicks to their remnants. Via momentum conservation, the other bodies in the interaction receive dynamical kicks in response. For simplicity, however, we ignore this subtlety when classifying the ejection mechanisms of specific escapers. Namely, whenever \texttt{CMC} removes an escaper immediately after a \texttt{fewbody} encounter that also involves SNe or mergers, we count only the direct SNe/merger remnants toward ejection via these secondary mechanisms. All other escapers from the \texttt{fewbody} interaction we attribute to the larger \texttt{fewbody} encounter itself. Our treatment is more precise for \textit{isolated} binaries outside of \texttt{fewbody}. Here, too, an SN may induce an opposite kick to the remnant's companion, ejecting it from the cluster. These escapers we \textit{do} count separately, properly attributing them to either their own SNe or the SNe of their (former) partners.

\section{Results} \label{S:escaper_properties}
We now examine the properties of escapers at $t_{\rm rmv}$, the time of removal from \texttt{CMC}, when they first satisfy Equation~(\ref{Eq:Giersz_energy_criterion}). This is prior to any further integration of their trajectories on their way out of the cluster and beyond (see the second paper in this series, Weatherford et al. 2023b, in preparation). Properties of interest include escape mechanism, location and velocity at $t_{\rm rmv}$, escaping star type, and escaping binary properties, such as semi-major axis, eccentricity, and mass ratio. First, however, we briefly demonstrate the overall cluster evolution.

\begin{figure*}[t!]
\centering
\includegraphics[scale=0.652]{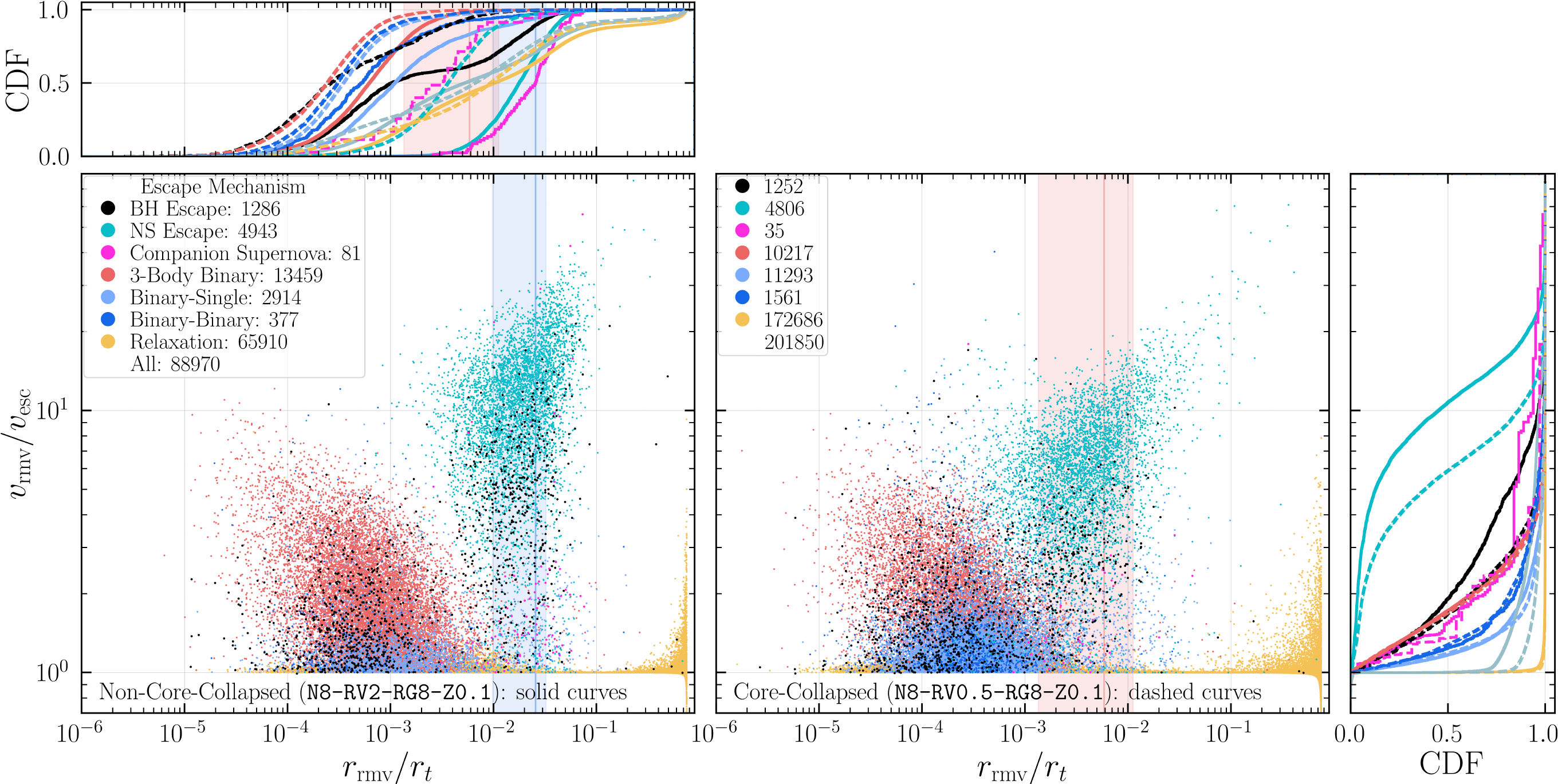}
\caption{The escapers from the archetypal NCC'd and CC'd cluster models (lower left and central panels, respectively), distributed according to their positions $r_{\rm rmv}/r_t$ and speeds $v_{\rm rmv}/v_{\rm esc}$ at the time of removal from \texttt{CMC}, $t_{\rm rmv}$. The top and right corner plots show the corresponding CDFs for $r_{\rm rmv}/r_t$ and $v_{\rm rmv}/v_{\rm esc}$, respectively, with solid (dashed) curves corresponding to the NCC'd (CC'd) model. In each panel, colors distinguish different escape mechanisms and the gray curves in the CDFs include all escapers. Regardless of escape mechanism, escapers (single or binary) containing a BH are shown in black while escapers containing an NS (but no BH) are shown in teal. All other escapers are categorized as follows: those caused by the induced kick from a binary companion's supernova (magenta), three-body binary formation (red), binary--single (light blue) and binary--binary (dark blue) strong encounters, and two-body relaxation (yellow). The legends also display the total number of escapers and the subtotals from each of the above categories. For each model, the vertical lines and surrounding shaded intervals indicate the median and 10th--90th percentile range of the theoretical density-weighted core radius $r_c(t_{\rm rmv})$ normalized by $r_t(t_{\rm rmv})$. Finally, note the escapers that appear at $r_{\rm rmv}/r_t\sim 1$ to have $v_{\rm rmv}/v_{\rm esc}<1$ are numerical artifacts. \texttt{CMC} \textit{does} require $v_{\rm rmv}\geq v_{\rm esc}$, but the recalculation of the escape energy from \texttt{CMC} output to make this plot in post-processing introduces enough numerical error to shift some escapers with $v_{\rm rmv}/v_{\rm esc}\approx 1$ to $v_{\rm rmv}/v_{\rm esc}$ just barely ${<}1$. The effect is small; the median $(v_{\rm esc} - v_{\rm rmv})$ for these escapers is only $2{\rm\ m\ s}^{-1}$ while only ${\approx}3\%$ of total escapers have $v_{\rm rmv}/v_{\rm esc}<0.999$.}
\label{fig:escape_mechanisms}
\end{figure*}

Figure~\ref{fig:Fig1} shows the time evolution of our GC simulations, distinguished in line style by Galactocentric distance $R_g$ and in color by virial radius $r_v$ (see legend). Recall that $R_g$ determines a cluster's tidal radius $r_t$, and $r_v$ its dynamical clock---Equation~(\ref{Eq:relaxation_timescale}); GCs with higher $R_g$ have larger tidal boundaries and those with lower $r_v$ (higher density) evolve faster. Together, $R_g$ and $r_v$ also determine how fully a GC fills its tidal boundary, quantifiable by the ratio $r_{99}/r_t$, where $r_{99}$ is the radius enclosing 99\% of the cluster mass. GCs born more tidally filling---i.e., with higher $r_v/R_g$ and $r_{99}/r_t$---have lower escape speeds in the halo and more easily evaporate. So while low $r_v$ generally hastens escape and cluster expansion (through dynamical heating), low $R_g$ GCs born at high enough $r_v$ can still rapidly disrupt due to their head start in filling their tidal boundaries. This duality is readily apparent in Figure~\ref{fig:Fig1}. The top two panels show the fractions retained of the initial number of particles $N(t)/N_i$ (left) and total cluster mass $M(t)/M_i$ (right). As expected, GCs with smaller $r_v$ typically evaporate faster for fixed $R_g$ while those with smaller $R_g$ evaporate faster for fixed $r_v$. But since the evaporation rate depends on both parameters, these individual trends are not always followed; either effect may outweigh the other in certain cases, e.g., for $R_g=2$~kpc and $r_v=4$~pc (solid yellow). Though the high $r_v$ implies slow relaxation, this model actually evaporates fastest since it is born the most tidally filling, evident from the lower right panel showing $r_{99}/r_t$.

Finally, the lower left panel shows a rolling average of the theoretical core radius $r_c$, expressed as a ratio to the half-mass radius $r_h$. As expected from their fast dynamics, the GCs with the smallest $r_v$ (0.5~pc; black) reach core collapse earliest---in fact within a Hubble time, demonstrable from their steep drops in $r_c/r_h$ between 8 and 13~Gyr. These drops accompany the transition from a centrally flat to a centrally steep surface brightness characteristic of an observationally CC'd state \cite[e.g.,][]{CMCCatalog,Kremer2021,Rui2021b}.

\subsection{Escapers by Escape Mechanism} \label{S:ejecta_by_escape_mechanisms}
To examine escape mechanisms in detail, we focus our attention on the \texttt{CMC} models most representative of typical NCC'd and CC'd MWGCs (models 2 and 8 from Table~\ref{table:model_properties}, respectively), as measured at a Hubble time under the definitions in Section~\ref{S:intro}. We record results for all models in the \hyperref[S:escape_conditions]{Appendix} (Tables \ref{table:escape_causes} and \ref{table:escaper_demographics}). Figure~\ref{fig:escape_mechanisms} shows the distribution of clustercentric position $r_{\rm rmv}/r_t$ and speed $v_{\rm rmv}/v_{\rm esc}$, as measured at the time of removal $t_{\rm rmv}$ from the NCC'd and CC'd GCs (lower left and central panels, respectively). Here, $v_{\rm esc}=\sqrt{2\alpha\phi(r_t)-2\phi(r_{\rm rmv})}$ is the escape speed, where $\phi(r)$ is the GC potential and $\alpha$ is the order unity constant defined in Equation~(\ref{Eq:Giersz_alpha}) to capture the effect of back-scattering. The upper and right corner plots show the corresponding cumulative distribution functions (CDFs) for $r_{\rm rmv}/r_t$ and $v_{\rm rmv}/v_{\rm esc}$ and different colors distinguish the escape mechanisms, as described in the caption and legends. In the corner plots, solid and dashed curves correspond to the NCC'd and CC'd GCs, respectively.

Several reassuring results are evident in Figure~\ref{fig:escape_mechanisms}. First, two-body relaxation (yellow) produces very low-speed escapers with $v_{\rm rmv}/v_{\rm esc}$ barely $>1$, except very near $r_{\rm rmv}/r_t \sim 1$. Here, $v_{\rm esc}$ is so small that weak encounters can produce $v_{\rm rmv}/v_{\rm esc}$ significantly greater than unity. Furthermore, about half of escapers from relaxation originate within the typical core radius at removal $r_c(t_{\rm rmv})$, indicated by the vertical line and shaded interval in each scatter plot. This echoes the understanding \citep[e.g.,][]{SpitzerShapiro1972} that stars near the escape energy in the cluster halo typically evaporate after first plunging back into the core, where the higher density makes relaxation much more efficient.

Also encouraging, the distribution of escapers containing a BH (black) is bimodal, reflecting the dominant two BH ejection scenarios; SNe kicks at modest radial positions of about one core radius and strong encounters in the deep core after formation and further mass segregation. Meanwhile, the overwhelming majority of escapers containing an NS (but no BH) are due to NS SNe kicks, which are stronger than BH SNe kicks, so are distributed at similar position but higher typical speed. Interestingly, the $r_{\rm rmv}$ distributions from both the BH and NS SNe both align very closely with each model's typical $r_c(t_{\rm rmv})$. This reflects that stars massive enough to form either type of compact object (${\approx}20\,M_\odot$ for NSs and ${\approx}40\,M_\odot$ for BHs) are proportionately much closer in mass to each other than to typical stars of ${\approx}0.5\,M_\odot$, so both mass segregate at roughly the same speed during their progenitors' very short lives. Continued mass segregation of the BHs to the cluster center and ensuing BH burning then forces retained NSs away from the deep core, leaving most of them at roughly the core radius.

More generally, strong encounters dominate escape from the deep core---though not from the core overall since two-body relaxation is so dominant (see, e.g., the escaper counts in the legends used to normalize the $r_{\rm rmv}/r_t$ CDFs at top). Strong encounters are especially dominant in the dense core of the CC'd GC (dashed), which also features several times more escapers from binary--single (light blue) and binary--binary (dark blue) interactions, as well as two-body relaxation (though not 3BBF in red, as we will discuss shortly). These reflect the increased density and correspondingly faster dynamics. The $v_{\rm rmv}/v_{\rm esc}$ CDFs from strong encounters (right) are similarly unsurprising; NS SNe kicks are by far the strongest, followed by ejections due to a binary companion's SN (magenta), 3BBF (since forming hard binaries releases much potential energy), and \texttt{fewbody} encounters. Though the difference in median seen in this CDF is small, binary--binary encounters tend to produce slightly higher $v_{\rm rmv}/v_{\rm esc}$ than binary--single encounters, especially in the distribution's high-$v$ tail. This reflects the additional binding energy available to exchange into post-encounter kinetic energy, whether through binary ionization, hardening under exchange, or even triple formation (though they are immediately broken by \texttt{CMC}).

\begin{figure*}[t!]
\centering
\includegraphics[scale=0.660]{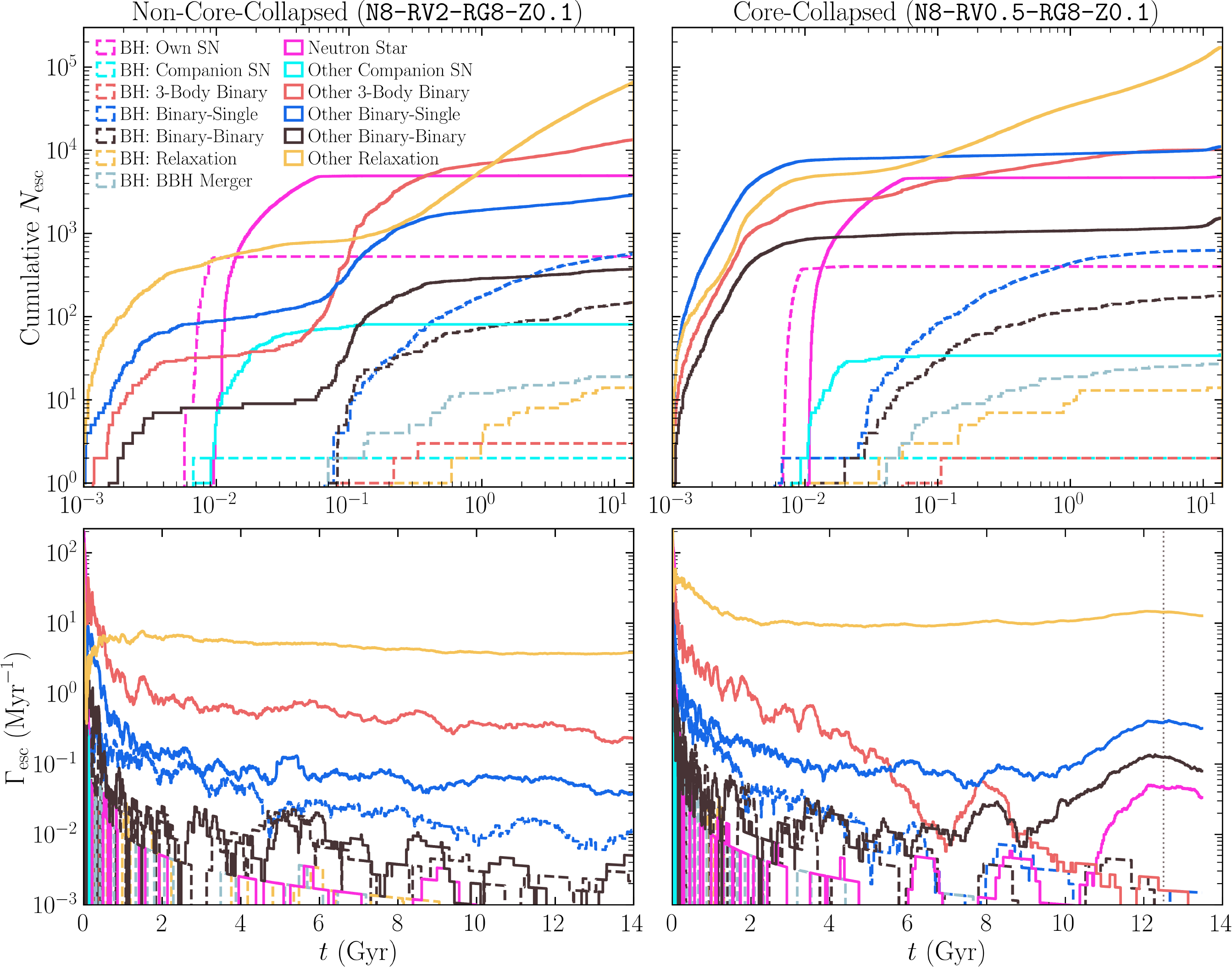}
\caption{Cumulative number of escapers (upper panels) and rolling average escape rate (lower panels) over time for the archetypal NCC'd and CC'd models (left and right panels, respectively). Note the time axis is logarithmic in the top row but linear in the bottom, to emphasize behavior near present. Escapers containing any BHs (dashed curves) are subdivided by escape mechanism: those ejected by SNe---whether their own (magenta) or a binary companion's (teal)---three-body binary formation (red), binary--single (blue) and binary--binary (black) encounters, binary BH mergers (gray), and two-body relaxation (yellow). Escapers containing any NSs (but no BHs) are shown in solid magenta regardless of escape mechanism (overwhelmingly SNe early on and cumulatively but mostly strong encounters during/after observable core collapse, indicated by the dashed vertical line; see Figure~\ref{fig:Fig1}). Typical stellar escapers (``other," containing no NSs or BHs) feature in the solid curves, with specific escape mechanisms colored as for the BH escapers. Finally, note the noise floor at ${\sim}10^{-3}{\rm\ Myr}^{-1}$ (from too small a sample size) in the lower panels decreases slightly as the rolling average window size increases linearly with time, from 1~Myr to 1.4~Gyr.}
\label{fig:escape_causes_time}
\end{figure*}

A more surprising result is that three-body binary formation contributes significantly to escape, \textit{especially} to the $v_{\rm rmv}$ distribution's high-speed tail. The latter is expected since 3BBF in \texttt{CMC} produces hard binaries, which exchange more potential energy into orbital speed. (Soft binaries are not considered since they quickly disrupt under weak encounters, anyway.) However, the overall prevalence of 3BBF is unexpected; this mechanism is typically considered a minor contributor to cluster evolution, its impact largely limited to BH binary formation \citep[e.g.,][]{Morscher2013,Morscher2015} and cessation of core collapse in clusters born without binaries---a largely academic scenario \citep[e.g.,][Ch.~23]{HeggieHut2003}. As we discuss later in this Section, this undervaluation derives largely from (now outdated) analytic arguments that neglect the extreme impact BHs have on 3BBF. For now, we note that Figure~\ref{fig:escape_mechanisms} suggests 3BBF is important to escape from GCs, due to its outsize influence on high-speed ejection. Notably, a few tens of 3BBF ejections of luminous stars---more than any other mechanism including companion SNe kicks---occur in the high hundreds of kilometers per second to even ${\gtrsim}10^3{\rm\ km\ s}^{-1}$. This makes 3BBF a possible contributor to \textit{hypervelocity} stars with speeds high enough to escape the MW \citep[for a review, see][]{Brown2015}. Yet while 3BBF dominates cumulative high-speed escape, it is not clear from Figure~\ref{fig:escape_mechanisms} alone that this dominance applies to present-day MWGCs.

\subsubsection{Time Evolution of Escape Mechanisms} \label{S:escape_mechanism_time_evolution}
To capture the crucial time dependence in the relative strengths of the various escape mechanisms, Figure~\ref{fig:escape_causes_time} shows the cumulative number of escapers $N_{\rm esc}$ (top) and escape rate $\Gamma_{\rm esc}$ (bottom) over time, classified by escape mechanism for the NCC'd (left) and CC'd (right) GCs. Mechanisms are separated for escapers containing any BHs (dashed curves) while those containing any NSs (but no BHs) are shown together (solid magenta). Typical stellar escapers, containing no NSs or BHs, are also separated (other solid curves).

Two-body relaxation dominates escape and this dominance is especially apparent at late times, i.e., the observable present for MWGCs. (However, note in Table~\ref{table:escape_causes} that exactly how much relaxation dominates over strong encounters depends on how tidally filling the GC is. As discussed earlier, tidally filling GCs at low $R_g$ can evaporate much more quickly from relaxation-coupled tides than GCs at high $R_g$ while the strong encounter rate changes little.) Relaxation (solid yellow) contributes ${\approx}94\%$ of $\Gamma_{\rm esc}$ at a Hubble time in the NCC'd GC, followed by 3BBF (solid red; $4\%$) and binary--single encounters (solid blue; $1\%$). There is negligible impact from other mechanisms, though some dominate at earlier times; a logarithmic scaling on the lower panels (or especially close look at the upper ones) would reveal that BH SNe (dashed magenta) dominate at $6\lesssim t/{\rm Myr}\lesssim 15$, NS SNe (solid magenta) at $15\lesssim t/{\rm Myr}\lesssim 90$, and other stars via 3BBF at $90\lesssim t/{\rm Myr}\lesssim 500$. Since most BHs escape by a Hubble time, we also see in the upper panels that binary--single strong encounters (dashed blue) and BH SNe each eject close to half of all BHs formed, with ${<}10\%$ ejected by binary--binary encounters and negligible impact from other mechanisms.

There are differences for the CC'd GC (but remember it is simply more dense, so dynamically older; the NCC'd GC would eventually collapse to a similar state if evolved beyond a Hubble time). First, due to the higher initial density, kicks via \texttt{fewbody} encounters, 3BBF, and relaxation contribute more to early $\Gamma_{\rm esc}$. So neither BH SNe nor 3BBF ever dominate and NS SNe dominate more weakly from 15--100~Myr. Cumulative $N_{\rm esc}$ is also several times higher at all times due to the shorter relaxation timescale. So, unlike ejecta from 3BBF or \texttt{fewbody} encounters, which escape at roughly the same rate as in the NCC'd GC by $t{\sim} 1$~Gyr, escape via relaxation remains faster. $\Gamma_{\rm esc}$ from \texttt{fewbody} encounters and relaxation also peaks once the GC reaches an observably CC'd state (vertical line; see Figure~\ref{fig:Fig1}). With few BHs (or NSs) left, the GC enters a phase in which WDs dominate central dynamics and support the core from further collapse via binary WD burning \citep[e.g.,][]{Kremer2021,Rui2021b,Vitral2022}. $\Gamma_{\rm esc}$ from these mechanisms then returns to gradually decreasing since $r_c(t)$ flattens out while $r_h$ continues to expand from binary burning. So while $r_c/r_h$ in Figure~\ref{fig:Fig1} continues contracting (much slower), the GC density and corresponding encounter rates actually decrease.

NS ejections also spike after core collapse, since the absence of BHs allows them to participate more vigorously in the denser central dynamics and eject themselves during \texttt{fewbody} interactions correspondingly more often. There are ${<}100$ NSs left in the CC'd GC by a Hubble time, so their heightened $\Gamma_{\rm esc}$ (about one NS every 20~Myr) may rapidly deplete their population on a gigayear timescale. However, this outcome neglects possible ongoing NS production through WD--WD mergers \citep{Kremer2021} and accretion-induced collapse. It is also sensitive to factors controlling the early NS population size, e.g., SN kicks and NS formation through giant collisions and tidal capture \citep[e.g.,][]{Ye2022}.

Finally, ejection via 3BBF (solid red) rapidly \textit{falls} during observable core collapse. This is highly counterintuitive since the total 3BBF rate in clusters is $\Gamma_{\rm 3bb}\sim n_s^3$ (or $\Gamma_{\rm 3bb,i}\sim n_s^2$ for any \textit{individual} body), where $n_s$ is the number density of singles. So $\Gamma_{\rm 3bb}$ scales even more steeply with $n_s$ than the total \texttt{fewbody} encounter rates---$(\Gamma_{\rm bs},\Gamma_{\rm bb})\sim(n_s n_b, n_b n_b)\sim n_s^2$, where $n_b$ is the binary number density---suggesting $\Gamma_{\rm 3bb}$ should also rise during core collapse. This intuition neglects steep dependence on the typical stellar mass $m$ and velocity dispersion $\sigma$: $\Gamma_{\rm 3bb,i}\propto n_s^2 m^5 \sigma^{-9}$---see Equation~(7.11) in \citet{BinneyTremaine2008}. $\Gamma_{\rm 3bb,i}$ in \texttt{CMC} follows the same scaling, with $m$ the mass of the new binary, and $n_s$ and $\sigma$ expressed as local averages---for more detail, see Section~2.3.1 of \citet{CMCRelease} but beware of a typo in their Equation~(23); the true \texttt{CMC} rate is 1/2 as large. Crucially, a GC prior to core collapse is in its binary BH burning phase, featuring a robust and dynamically dominant central population of BHs. These BHs are typically $10$--$20$ times more massive than other bodies in the core, so their dominance of central dynamics, including 3BBF, enhances $\Gamma_{\rm 3bb}$ by a factor of ${\sim}10^6$ over a first-order approximation based on average \textit{stellar} mass \citep[e.g.,][]{BinneyTremaine2008}. Upon the loss of BHs leading to core collapse, the 3BBF ejection rate craters and even the higher post-collapse density cannot raise it enough to match the earlier BH-driven rate, at least until dynamics eject most remaining \textit{stellar} binaries now responsible for supporting the core.

\begin{figure*}[t!]
\centering
\includegraphics[scale=0.661]{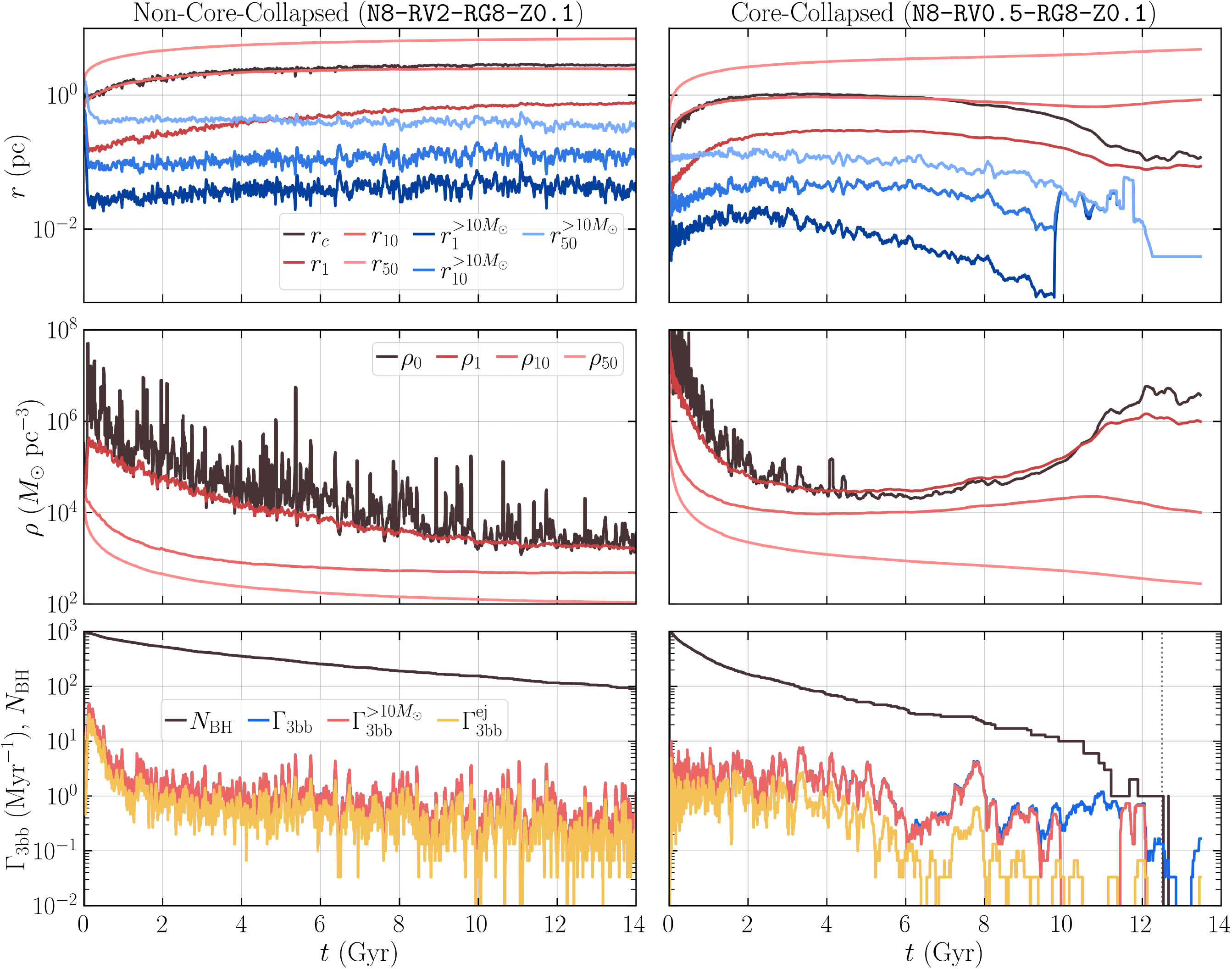}
\caption{For the archetypal NCC'd and CC'd models (left and right columns, respectively) the upper panels show time-rolling averages of the density-weighted core radius $r_c$ \cite[black; from][]{CasertanoHut1985} and Lagrange radii containing (1\%, 10\%, 50\%) of the cluster mass, counting all cluster objects (reds; $r_{1}$, $r_{10}$, $r_{50}$) and just those with masses $10 < m/M_\odot \leq 100$ (blues; $r_{1}^{>10M_\odot}$, $r_{10}^{>10M_\odot}$, $r_{50}^{>10M_\odot}$). Central panels: time-rolling averages of the density-weighted central density $\rho_0$ \cite[black; also from][]{CasertanoHut1985} and average densities within each the above Lagrange radii counting all cluster objects (reds; $\rho_{1}$, $\rho_{10}$, $\rho_{50}$). Lower panels: time-rolling averages of the total number of BHs in the cluster $N_{\rm BH}$ (black), the total 3BBF rate $\Gamma_{\rm 3bb}$ (blue), the 3BBF rate involving at least one body with mass $m>10\,M_\odot$ ($\Gamma_{\rm 3bb}^{>10\,M_\odot}$; red), and the ejection rate via 3BBF $\Gamma_{\rm 3bb}^{\rm ej}$ (yellow). Note that $\Gamma_{\rm 3bb}^{>10\,M_\odot}$ often obscures $\Gamma_{\rm 3bb}$ due to their near equality at most times. The background grid lines in gray help guide the eye in noticing that many of the peaks/troughs in the 3BBF rates correspond closely in time to peaks/troughs in density (troughs/peaks in radius), yet the relative magnitudes are inconsistent; i.e., large density peaks (radius troughs) do not necessarily lead to large peaks in the 3BBF rates. Finally, the rolling average window size is 30~Myr in all panels.}
\label{fig:3bb}
\end{figure*}

\subsection{A Closer Look at Three-body Binary Formation} \label{S:closer_look_at_3bb}
The above reasoning is sound; BH mass should increase 3BBF. Yet it still seems to contradict more subtly the findings of \cite{Morscher2013,Morscher2015} that most 3BBF occurs during transient collapses of the central BH population (gravothermal oscillations, \textit{not} observable core collapse). Since BHs are so massive and compact, these transient BH collapses can be significantly deeper than those achieved by stellar bodies (primarily central WDs) after observable core collapse. In other words, the high 3BBF rate during the BH burning phase may not be due to the mass of BHs but rather their capacity to reach extreme densities. Critically, however, \cite{Morscher2013,Morscher2015} only allowed 3BBF to occur between three BHs, greatly limiting $\Gamma_{\rm 3bb}$ overall (e.g., between one BH and two stars). Due to BHs' relative rarity, it is not surprising that 3BBF so constrained would only be efficient during transient collapses of the BH population. To show that our high 3BBF ejection rate $\Gamma_{\rm 3bb}^{\rm ej}$ during the BH burning phase is truly due to the mass of BHs, rather than their ability to collapse to extreme density, we must look more closely at how $\Gamma_{\rm 3bb}$ and $\Gamma_{\rm 3bb}^{\rm ej}$ change with density.

The top row of Figure~\ref{fig:3bb} shows the time evolution of the theoretical core radius $r_c$ and two sets of Lagrange radii enclosing (1\%, 10\%, 50\%) of the cluster mass, counting all cluster objects and just those with masses from $10$--$100\,M_\odot$. Again the NCC'd and CC'd GCs are shown at left and right, respectively. The center row shows the central density and average densities within the above Lagrange radii for all cluster bodies while the bottom row shows the number of BHs in the cluster $N_{\rm BH}$, and 3BBF rates: overall ($\Gamma_{\rm 3bb}$), involving at least one body more massive than $10\,M_\odot$ ($\Gamma_{\rm 3bb}^{>10M_\odot}$), and leading to ejection ($\Gamma_{\rm 3bb}^{\rm ej}$). Though all curves have been smoothed for visibility using a rolling average with a 30 Myr window size, gravothermal oscillations are still easily apparent, with numerous spikes in central density and smaller troughs in the Lagrange radii limited to high masses.

As observed by \cite{Morscher2013}, the density peaks (radii troughs) typically occur simultaneously with peaks (troughs) in $\Gamma_{\rm 3bb}$ (blue; though note this is typically obscured by the nearly equal $\Gamma_{\rm 3bb}^{>10M_\odot}$ in red). However, the relative magnitude of these peaks/troughs is not especially consistent; even sharp density spikes often lead to relatively shallow peaks in 3BBF. Thus, much of the 3BBF occurs outside of the BH collapse phases. More tellingly, the CC'd GC features much higher \textit{typical} densities (lower characteristic radii) than the NCC'd GC, due both to its smaller initial $r_v$ and its early core collapse. This remains true of the Lagrange densities of the bodies more massive than $10\,M_\odot$---which are all BHs after the first tens of megayears---and especially so by a Hubble time. Yet even at this point, $\Gamma_{\rm 3bb}$ remains at least a few times higher than in the CC'd GC. So while gravothermal oscillations help induce 3BBF, they are \textit{not} the primary cause of 3BBF overall. $\Gamma_{\rm 3bb}$ instead follows very closely the evolution of $N_{\rm BH}$ (black), leaving BH mass enhancement as the primary cause of 3BBF in our models.

Figure~\ref{fig:3bb} also reveals a couple further interesting nuances to escape via 3BBF. First, $\Gamma_{\rm 3bb}^{>10M_\odot}$ is practically indistinguishable from $\Gamma_{\rm 3bb}$, at least prior to core collapse, demonstrating that almost all 3BBF in the NCC'd GC---and correspondingly, almost all 3BBF-induced ejections---involves at least one BH. Secondly, Figure~\ref{fig:3bb}'s lower right panel demonstrates that $\Gamma_{\rm 3bb}^{\rm ej}$ decreases more rapidly with BH loss than $\Gamma_{\rm 3bb}$. So $\Gamma_{\rm 3bb}^{\rm ej}$ is even \textit{more} sensitive to component mass than $\Gamma_{\rm 3bb}$. This is due to momentum conservation in the center of mass (COM) frame of the three-body interaction, which kicks the binary and single in inverse proportion to their masses. Binaries formed via 3BBF involving a BH will almost always contain the BH (as the most massive object), so these binaries are significantly more massive than those formed in the absence of BHs. They then receive a smaller kick, while the single picks up a larger kick, making its ejection more likely.

\begin{figure}[t!]
\centering
\includegraphics[scale=0.592]{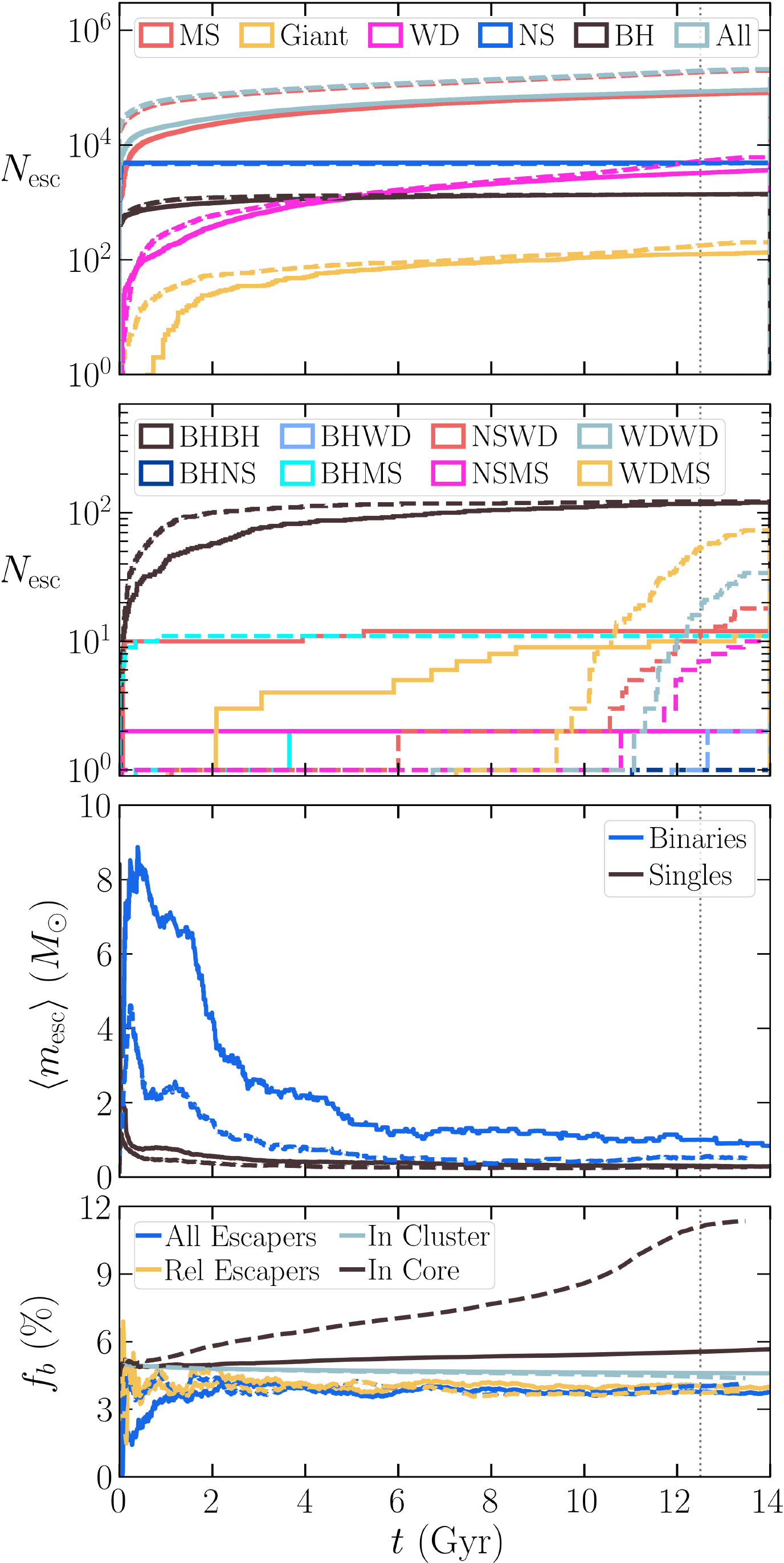}
\caption{Top two panels: cumulative number of escapers over time for the archetypal NCC'd and CC'd models (solid and dashed curves, respectively). The top panel tracks all escaping bodies by stellar type (binary components count individually); MS = main-sequence, WD = white dwarf, NS = neutron star, BH = black hole, and ``Giant" includes stars in the Hertzsprung gap. The second panel tracks binaries containing at least one compact object. Note there are no such ejecta with Giant companions. Third panel: Time-rolling average mass of escaping bodies. Bottom panel: time-rolling average binary fractions in the cluster overall (gray), within the cluster's density-weighted core radius (black), among all escapers (blue), and among stellar relaxation-induced escapers only (yellow). In all panels, the vertical line indicates the onset of WD binary burning (observational core collapse; see Figure~\ref{fig:Fig1}).}
\label{fig:escaper_demographics}
\end{figure}

\subsection{Escaper Demographics} \label{S:escaper_demographics}
We now examine escaper demographics. For the same two archetypal GCs, the upper three panels of Figure~\ref{fig:escaper_demographics} show the cumulative number of escapers over time for different stellar/binary types. Consistent with our earlier look at escape mechanisms, the top panel shows that in the NCC'd cluster (solid curves), NSs (blue) briefly surpass MS stars (red) as the dominant population from ${\sim}10$--$100$~Myr. At all other times, the MS dominates escape. Due to the CC'd GC's higher density and faster relaxation, more MS stars escape at early times via both strong encounters and relaxation, so NSs never dominate.

The next panel tracks ejection of binaries containing at least one compact object (BH, NS, or WD). Except for BH--BH pairings, the NCC'd GC strongly disfavors ejection of any such binaries. This results from the low initial density and correspondingly low encounter rates, exacerbated by long BH retention; the associated binary BH burning mostly excludes lighter compact objects from the central dynamics. BH--BH binary ejection is only slightly faster in the CC'd GC due to the higher density, but the ejection rates of the other binary species are quite different from the NCC'd case. BH loss precipitating core collapse allows NSs and WDs to participate strongly in central dynamics; tens of ejections each of NS--WD and WD--WD binaries, plus ${\sim}10$ NS--MS and nearly 100 WD--MS binaries, occur after the steepest phase of core contraction around 10~Gyr (see Figure~\ref{fig:Fig1}'s dashed black curve).\footnote{Curiously, ${\sim}10$ each of NS--WD and WD--MS binary ejections occur far earlier in the NCC'd GC, a result that should be viewed skeptically. These few are almost all primordial binaries and feature initial mass ratios near unity and significant mass transfer (especially the NS--WDs). Their prompt ejection from the NCC'd model but not the denser CC'd model likely arises from a \texttt{CMC} prescription that scales the initial semi-major axis linearly with $r_v$. This causes the CC'd GCs to start with harder binaries, which may then be more susceptible to mass transfer instabilities in stellar evolution with \texttt{BSE}.}

\begin{figure*}[t!]
\centering
\includegraphics[scale=0.645]{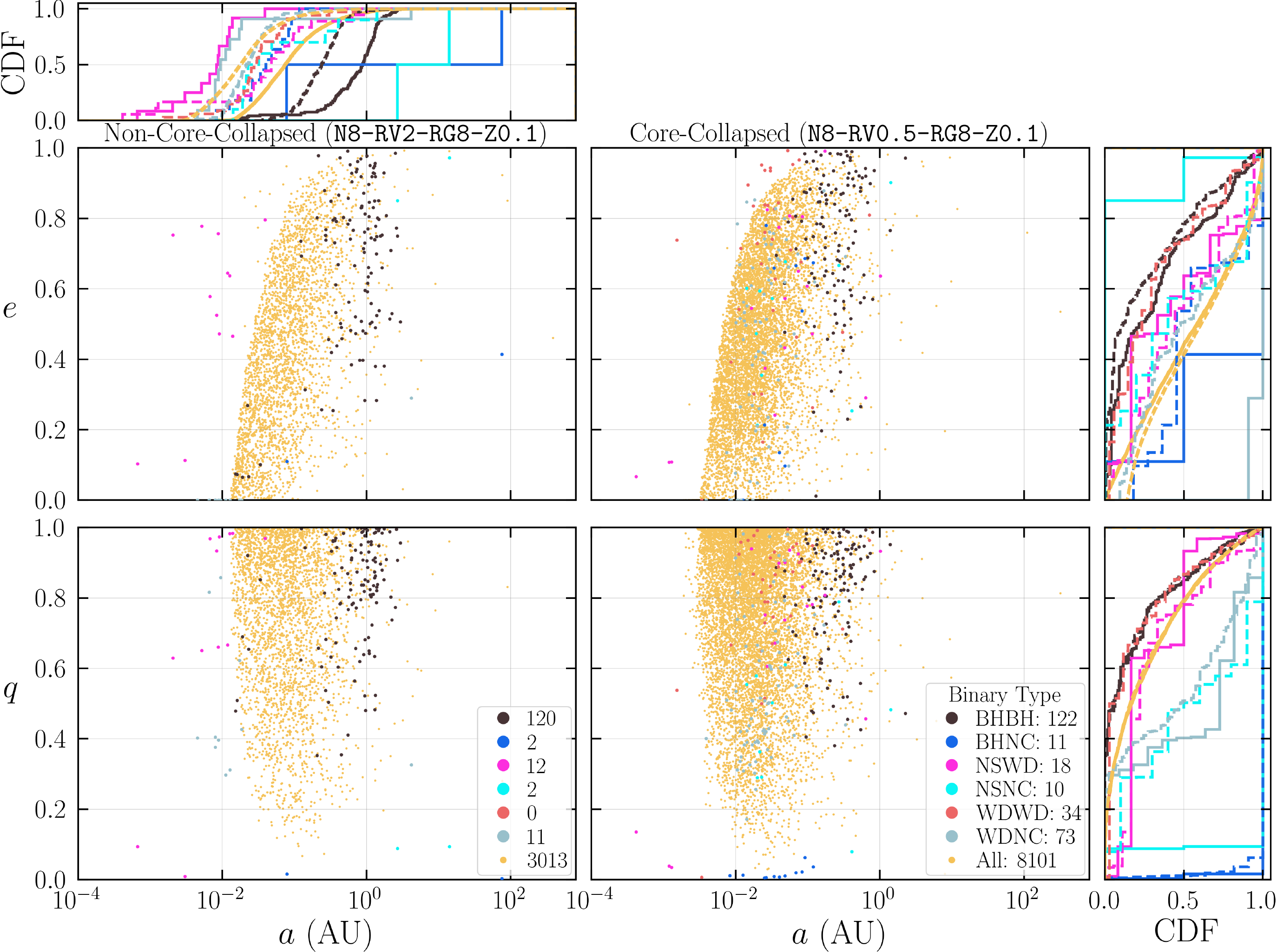}
\caption{Properties of escaping binaries from the archetypal NCC'd and CC'd models (left and central scatter plots, respectively), distributed according to their eccentricity $e$, mass ratio $q=m_{\rm secondary}/m_{\rm primary}$, and semi-major axis $a$, all at the time of removal $t_{\rm rmv}$ from \texttt{CMC}. In clockwise order, the top and rightmost corner plots show the corresponding CDFs for $a$, $e$, and $q$, respectively, with solid (dashed) curves corresponding to the NCC'd (CC'd) model. In each panel, colors distinguish several types of binaries: BH--BH (black), BH--NC (blue), NS--WD (magenta), NS--NC (teal), WD--WD (red), WD--NC (gray), and all binaries (yellow). Here, ``NC" indicates when one of the binary components is \textit{non-compact}, i.e., any body (overwhelmingly MS stars) other than a WD, NS, or BH. For each model, the legends in the lower panels also display the total number of binary escapers and the subtotals from each of the above categories.}
\label{fig:binary_properties}
\end{figure*}

The third panel shows the time-rolling average mass of escaping binaries (blue) and singles (black). There is a large peak for binaries in the first couple gigayears due to frequent ejection of BH binaries through hardening encounters with other bodies. This peak is significantly smaller for the core-collapsed cluster because several times more binaries overall are ejected via faster dynamics, while the number of BH binary ejections is roughly the same due to their limited supply (see the second panel). In both clusters by a Hubble time, the average mass of single ejections is ${\approx}0.3\,M_\odot$ (binary masses ${\approx}0.5\,M_\odot$ in the CC'd GC and ${\approx}0.8\,M_\odot$ in the NCC'd GC). This is about $50\%$ less than the average stellar mass in the cluster, reflecting the preferential ejection of low-mass stars due to their lower inertia.

The lowest panel compares time-rolling averages of the binary fractions in the cluster overall (gray), within its theoretical core radius (black), among all escapers (blue), and among stellar relaxation-induced escapers only (yellow). The overall cluster binary fraction $f_b$ is nearly constant, decreasing only slightly over a Hubble time due to dynamical destruction of binaries through strong encounters and collisions \citep[e.g.,][]{Hurley2007,Chatterjee2010}. In contrast, the binary fraction in the core ($f_{b,c}$) increases due to mass segregation, and does so much faster in the CC'd GC \citep[see][]{Chatterjee2010}, due to its faster relaxation. Since relaxation-induced escape primarily takes place in the core (e.g., Figure~\ref{fig:escape_mechanisms}), one might expect the binary fraction among \textit{only} relaxation-induced escapers ($f_{b,{\rm relesc}}$) to trace $f_{b,c}$---or at least somewhere in between $f_{b,c}$ and $f_{b}$ overall. Yet $f_{b,{\rm relesc}}$ is nearly identical in both GCs and noticeably lower than $f_{b,c}$ or $f_{b}$. This reflects the higher mass of binaries relative to singles; momentum conservation during weak encounters with singles then causes binaries to receive smaller kicks. This may damp the escaper binary fraction more in the CC'd GC since the faster dynamics allows for more binary exchange interactions, which tend to increase component mass. Finally, due to the dominance of relaxation-induced escape, it is unsurprising that the \textit{overall} escaper binary fraction $f_{b,{\rm esc}}$ closely tracks $f_{b,{\rm relesc}}$ from relaxation only. Less obviously, $f_{b,{\rm esc}}$/$f_{b,{\rm relesc}}$ is slightly higher in the CC'd GC since its higher density generates more strong \texttt{fewbody} encounters, which feature higher kicks.

\subsection{Escaping Binaries} \label{S:escaping_binaries}
More detailed properties of binary escapers merit further examination. Figure~\ref{fig:binary_properties} shows the distributions in semi-major axis ($a$; CDF at upper left), eccentricity ($e$; CDF at center right), and secondary-to-primary mass ratio ($q$; CDF at lower right) among all escapers from the NCC'd and CC'd GCs (solid and dashed curves, respectively). The four scatter plots show the two-dimensional shape of these distributions for each cluster (NCC'd in the lower left pair of panels and CC'd in the central pair). Different colors correspond to different binary types, as described in the caption. It is readily apparent in the top panel that binaries escaping from the CC'd GC have smaller $a$ than binaries in the NCC'd GC, due to faster dynamics (faster hardening) and the higher central escape speed (allowing for more hardening before ejection).

In both GCs, the eccentricities of ejected BH--BH binaries (black) are nearly thermal, as expected for binaries in statistical equilibrium \citep[e.g.,][]{Jeans1919,Ambartsumian1937,Heggie1975}. Meanwhile, ejected binaries overall (yellow) strongly disfavor high $e$---as do in-cluster binaries at late times, not shown---consistent with surveys of the Galactic field \citep[for recent reviews, see, e.g.,][]{DucheneKraus2013,MoeDiStefano2017}. Since many field stars likely escaped from disrupted star clusters \citep[e.g.,][]{LadaLada2003}, the result that stellar escapers from \texttt{CMC} have a similarly flat $e$ distribution is encouraging. However, it is also somewhat counterintuitive theoretically, since dynamically formed binaries tend to have thermal eccentricities \citep{Heggie1975} and \texttt{CMC} correspondingly draws $e$ from a thermal distribution for both primordial binaries \textit{and} those generated via 3BBF.

This tension echoes the finding of \cite{Geller2019} that clusters---simulated with both \texttt{CMC} and direct $N$-body codes---cannot thermalize initially uniform/near-uniform $e$ distributions in MS stars, even after several Hubble times. (In our case, the tension is worse because the primordial binaries start thermal and actively become less so for both in-cluster and ejected binaries.) This counterintuitive result arises from a couple factors. First, while frequent strong scattering encounters can increase $e$, high $e$ also translates to shorter pericenter distances and correspondingly more stellar collisions, destroying binaries that grow too eccentric. Less extremely, small pericenter can lead to mass transfer, common envelope, and strong tides in binary stellar evolution. As dissipative processes, these circularize $e$. Hard binaries dynamically ejected from the dense core can be subject to these processes for extended periods with few interruptions from $e$-enhancing encounters.

The clusters' binary mass ratios $q=m_{\rm secondary}/m_{\rm primary}$ are also similar. Escaping binaries overall exhibit a strong tendency toward $q\sim 1$, especially for BH--BH binaries, though some pairings intuitively exhibit $q\ll 1$---e.g., BH--NC binaries (blue) between a BH and any non-compact object (overwhelmingly MS stars). In fact, the overall $q$ distribution for escapers is very similar to that for in-cluster binaries at a Hubble time (not shown), with only a slightly higher skew toward $q\sim 1$. So the $q$ distribution predominantly reflects strong \texttt{fewbody} dynamics in GCs, in which binaries preferentially swap in companions of higher mass \citep[e.g.,][Ch.~19]{HeggieHut2003}. While the most massive objects in old GCs are BHs, followed distantly by NSs and massive WDs, the upper end of the MS remains by far the most numerous massive population, so binaries in GCs generally favor MS--MS binaries with both component masses near the MS turnoff, i.e., a mass ratio $q\sim1$. This is qualitatively consistent with binaries surveyed in MWGCs, characterized by a roughly uniform mass ratio for $q\lesssim 0.9$ with a sharp over-density for $q\gtrsim 0.9$ \citep{Milone2012}. In contrast, binaries in the Galactic field generally exhibit more asymmetric mass ratios \citep[e.g.,][]{MoeDiStefano2017}. So the tendency toward $q\sim 1$ in binaries escaping from GCs could help in identifying field binaries with origins in nearby MWGCs. However, this should only be leveraged as a supplement to more robust signals, e.g., proper motion, position, and metallicity.

Finally, we find that even in the CC'd GC, most binaries (${>}80\%$) escape via two-body relaxation, with binary--binary strong encounters contributing most of the rest. Binary--single encounters eject only a small fraction of binary escapers (${\lesssim}5\%$ each). In \texttt{CMC}, 3BBF almost never ejects the binary (twice throughout all 12 simulations analyzed). This occurs since 3BBF is by far more efficient when a BH (massive body) is involved. Consequently, the typical binary-to-single mass ratio at the end of a 3BBF encounter in \texttt{CMC} is ${\sim}10$. To conserve momentum in the encounter's center of mass frame, the binary thus receives a much smaller recoil kick than the single.

\section{Discussion} \label{S:disc}
Before summarizing our findings and future work, we now discuss the surprising impact of three-body binary formation on escape from GCs and limitations to our analysis.

\subsection{Relevance of Three-body Binary Formation} \label{S:3BBF}
We find three-body binary formation occurs often enough in GCs to cumulatively power many high-speed ejections in the NCC'd GCs typical of the MW but \textit{not} GCs that are observably CC'd (i.e., with steep central surface brightness). This finding completely reverses common arguments that 3BBF is negligible until \textit{after} core collapse, especially given similar doubts about its dynamical impact even then. These arguments rest on flawed assumptions while the apparent reversal in behavior relative to \textit{core collapse} stems from evolution in usage of the term.

First, analytic estimates of the 3BBF rate \citep[e.g.,][]{Heggie1975,GoodmanHut1993,BinneyTremaine2008} have sometimes been misinterpreted, in conjunction with a low assumed stellar mass, to suggest the rate is negligible prior to core collapse \citep[e.g.,][]{Hut1985,FreitagBenz2001,Joshi2001}, or even thereafter \citep[e.g.,][]{Statler1987,Hut1992}. Such studies generally predate the cluster modeling community's widespread incorporation of primordial binaries and realistic IMFs---and thereby the modern consensus that GCs retain robust BH populations prior to collapse. As noted in Section~\ref{S:ejecta_by_escape_mechanisms}, BHs greatly enhance 3BBF due to the 3BBF rate's steep mass dependence \citep[see also][]{Kulkarni1993,OLeary2006,Banerjee2010,Morscher2013,Morscher2015}. Without such massive bodies, efficient 3BBF would indeed require extreme cluster density only achieved in deep core collapse. Additionally, binary burning prevents modern GC models with primordial binaries from reaching such extremes \citep[e.g.,][]{GoodmanHut1989}. Modern usage of the term \textit{core collapse} has changed accordingly and is not limited to such deep collapse halted by 3BBF. Rather, as in this study, it often refers to the core contraction arising from BH ejections, which transition the core from binary BH burning to weaker binary WD burning, as described in Section~\ref{S:intro}. The lower mass of WDs causes the 3BBF ejection rate to drop precipitously due to the overall 3BBF rate's steep mass dependence, made even steeper by the extra mass dependence in the leftover single's speed (via momentum conservation), as discussed in Section~\ref{S:closer_look_at_3bb}. This overwhelms the boost from higher density post-collapse. Only once strong encounters harden and eject the remaining binaries can deeper collapse occur. The combination of the above factors---neglect of BH populations and updated terminology---explain the apparent reversal in 3BBF's behavior in relation to core collapse.

Second, binaries formed via 3BBF that are especially soft or especially hard do not survive long enough to contribute much to binary burning. In the former case, strong encounters quickly disrupt the binary, while in the latter they harden it until the components eventually merge, if they are not ejected from the cluster first \citep[e.g.,][]{HutInagaki1985,McMillan1986,GoodmanHernquist1991,Bacon1996,ChernoffHuang1996,Fregeau2004}. 
This reasoning has previously been used to justify neglecting 3BBF in \texttt{CMC}, either entirely \citep[e.g.,][]{Joshi2000,Fregeau2003} or if any of the bodies is not a BH \citep[e.g.,][]{Morscher2015}. Unfortunately, this neglects formation of binaries more intermediate in hardness that survive long enough to contribute substantially to binary burning, as well as 3BBF of BHs with non-BHs. Binary survival is also entirely irrelevant to 3BBF's impact on the escape process; formation of even a short-lived binary in a three-body encounter still kicks the third body. 3BBF can therefore contribute significantly to cluster dynamics via both binary burning and ejection.

Our findings are susceptible to inaccuracies, too, however, especially since \texttt{CMC} uses an approximate recipe \citep[][and references therein]{CMCRelease} for 3BBF rather than direct integration with, e.g., \texttt{fewbody}. In brief, \texttt{CMC} divides a radially sorted list of all singles in the GC into sets of three and computes the 3BBF probability for a binary with a certain minimum hardness based on the three masses and local average relative velocity and number density. The prescription in \texttt{CMC} automatically pairs two most massive bodies in a set to form a binary. Realistically, this may be the most likely pairing but is not guaranteed. This skews the mean mass ratio between the binary and leftover single to be $m_b/m_s\sim 10$, making immediate ejection of the binary from the GC exceptionally difficult as the ratio of the final speeds in the 3BB interaction's center of mass (COM) frame is $v_b/v_s = m_s/m_b\sim 0.1$. This sensitivity to the mass ratio means automatic pairing of the two most massive bodies may bias the ejection speeds.

\texttt{CMC}'s 3BBF recipe also samples eccentricity $e$ from a thermal distribution for all values of semi-major axis $a$, but scattering experiments show $\langle e \rangle$ increases with $a$ and is higher than thermal even at the hardest $a$ tested \citep[1/2 the hard-soft boundary;][]{AarsethHeggie1976}. \texttt{CMC} further assumes the probability $P$ of binary formation is 100\% when the three bodies all pass within a region of size $a$, though the tests show $P\approx54\%$ for $a$ at the hard-soft boundary. Notably, \cite{AarsethHeggie1976} only studied the case of equal masses. So while \cite{Morscher2013} found good agreement between 3BBF rates from \texttt{CMC} and direct $N$-body methods, the above uncertainties in light of 3BBF's dominance of high-speed ejections suggest that upgrading the 3BBF recipe to use \texttt{fewbody} may be highly worthwhile.

\subsection{Additional Limitations} \label{S:limitations}
Our analysis is subject to several further limitations in addition to those affecting three-body binary formation. Some were previously noted in Section~\ref{S:escape_mechanisms_in_CMC}, but we recap them here with an emphasis on how they may affect the results, as well as options to address them in the future.

First, and most generally important to escape physics, \texttt{CMC} assumes spherical symmetry and therefore is not reliable close to the tidal boundary. Fortunately, however, even two-body relaxation mostly ejects bodies from the cluster core, where the true tidal potential is still very spherical (e.g., Section~\ref{S:ejecta_by_escape_mechanisms} and \citealt{SpitzerShapiro1972}). So spherical symmetry matters less given our focus on escape mechanisms rather than escape trajectories. (Direct integration of escaper trajectories from the point they first meet the energy criterion mitigates this limitation in follow-up work.)

Second, \texttt{CMC} neglects time-dependent external tides from eccentric or disk-crossing GC orbits, including disk and bulge shocking, which would cause the evaporation rate in Figure~\ref{fig:escape_causes_time} to fluctuate and increase overall. The Monte Carlo cluster modeling method can accommodate tidal shock heating\citep[e.g.,][]{SpitzerChevalier1973,SpitzerShull1975}. However, the method's faster H\'{e}non-style variant used in most modern implementations (including \texttt{CMC}) is not ideally suited for this, being an orbit-averaged approach; rather than evolving each body's orbit on the crossing timescale, it averages over the effects on the (much longer) relaxation timescale. Since tidal shocks occur on the crossing timescale, codes like \texttt{CMC} can only approximate these processes. Yet work on implementing such prescriptions into \texttt{CMC} is in progress \citep[see also discussion in][]{Rodriguez2022a}.

Third, while \texttt{CMC} can and does form IMBHs of several hundred solar masses in certain circumstances \citep[see, e.g.,][]{Gonzalez2021,Weatherford2021,Gonzalez_2022}, the code cannot yet fully model the influence of an off-center IMBH in the range $10^3$--$10^4\,M_\odot$. Efforts are underway to add this capability, but for the time being, \texttt{CMC} cannot explore the dynamical effects of truly massive BHs in GCs. As discussed in Section~\ref{S:IMBHs}, the presence of such a massive body in any GC could significantly increase the rate of high-speed ejections from the cluster. Hence, future analysis of IMBH-induced ejections in full GC models---whether direct $N$-body or Monte Carlo codes---may help constrain the possible presence of IMBHs in GCs with well-measured ejection speed distributions \citep[e.g.,][]{SubrFragione2019}.

On a lesser note, \texttt{CMC} does not consider strong two-body encounters, which may slightly increase the ejection rate at intermediate speeds up to several times the local escape speed. \texttt{CMC} can be upgraded to do so by passing some two-body encounters to \texttt{fewbody} instead of the main relaxation algorithm; while the overall effect of such a change is likely minimal due to the rarity of such encounters, it may be worthwhile given the relative ease of such an upgrade. It is also a direct first step to a more complex (and likely more impactful) \texttt{fewbody}-based implementation of three-body binary formation.

Finally, \texttt{CMC} does not yet feature asymmetric mass ejection during physical collisions, tidal capture, or tidal disruption events. Consideration of kicks due to asymmetric mass loss during these phenomena may have important effects on the upper end of the ejection speed distribution and the presence of exotica, such as stripped stars in the outskirts of GCs. Note this includes tidal tails and stellar streams---which primarily channel low-speed escapers---since kicks due to asymmetric mass loss from these events may be quite modest (Section~\ref{S:near_contact_recoil}). These considerations may be more important for observably CC'd GCs, given their higher collision and tidal disruption rates. The severe shortage of BHs in these GCs would also heighten the impact of collisions and disruptions involving NSs/WDs at the expense of those involving BHs. For these reasons, improvement of \texttt{CMC}'s collision and tidal disruption prescriptions, grounded in hydrodynamical simulations, is an ongoing priority \citep[e.g.,][]{Kiroglu2022b,Kremer2022_BH,Kremer2022_NS,Ye2022}.

\section{Summary and Future Work} \label{S:summary}
The Gaia telescope has revealed numerous stellar streams and traced the origin of several to specific MWGCs. This connection and the streams' importance to Galactic archeology highlight the need for further examination of escape from GCs. As a first step toward a detailed comparison of the ejecta from our \texttt{CMC} models to extra-tidal stellar populations from the Gaia survey in the vicinity of MWGCs, we have studied escape mechanisms in \texttt{CMC}. Consistent with long-standing theory \citep[e.g.,][]{SpitzerShapiro1972} and numerical modeling \citep[e.g.,][]{PeretsSubr2012,MoyanoLoyolaHurley2013}, we find that two-body relaxation in the cluster core dominates the overall escape rate while central strong encounters involving binaries contribute especially high-speed ejections, as do SNe and GW-driven mergers. We also find the escape rate at a Hubble time in observably CC'd clusters reflects the transition from binary BH burning to binary WD burning \citep{Kremer2021}, boosting late ejections of WD and NS binaries.

We have also shown for the first time that three-body binary formation plays a significant role in the escape dynamics of NCC'd GCs typical of those in the MW. BHs are an essential catalyst for this process due to the 3BBF rate's sensitive dependence on binary mass. As long as a significant BH population remains in the cluster's core, 3BBF dominates the rate of present-day high-speed ejections over any other mechanism, including standard binary--single and binary--binary scattering interactions. This includes production of hypervelocity stars with speeds in the hundreds of kilometers per second to even ${\gtrsim}10^3{\rm\ km\ s}^{-1}$. 3BBF then plummets with the loss of BHs at the onset of observable core collapse.

Except for BH--BH binaries, we find that binary escapers from GCs (as well as in-cluster binaries at late times) are far more circular than expected from a thermal distribution, consistent with observations of the Galactic field \citep[e.g.,][]{MoeDiStefano2017}. This occurs even though \texttt{CMC}'s initial binary eccentricities are thermal, including for those produced via 3BBF. This echoes \cite{Geller2019}, who found that since dynamical hardening and collisions deplete eccentric binaries, realistic cluster dynamics do not necessarily thermalize initially uniform eccentricities, contrary to arguments based on thermal equilibrium. Our findings go a step further by demonstrating that dynamics actively de-thermalize the eccentricity distribution, which motivates drawing initial eccentricities in future modeling from flatter distributions more consistent with observations.

Finally, while this study provides a broad sense of the escape mechanisms and demographics of escapers from GCs, the results are not immediately comparable to Gaia observations. In our next work (Weatherford et al. 2023b, in preparation), we therefore integrate the trajectories of \texttt{CMC} escapers in a full Galactic potential and continue their internal stellar evolution to construct realistic velocity distributions in the extra-tidal regions of \texttt{CMC} models and mock surface brightness profiles (in the Gaia bands) extending out to several tidal radii. We will explore how well this post-processing approach reproduces tidal tails, currently unclear due to \texttt{CMC}'s assumed spherical symmetry. In later work (Weatherford et al. 2023c, in preparation), we will identify likely past members (\textit{extra-tidal candidates}) of specific MWGCs and directly compare the mock ejecta from our cluster models to the Gaia data. As shown recently by \cite{Grondin2023}, combining precise astrometry with chemical tagging is an especially promising method of identifying such extra-tidal ejecta, and may even be used to identify stars from particular ejection mechanisms. Ultimately, we hope to better understand stellar stream formation and, in an ideal case, leverage the new observables from Gaia to better constrain uncertain properties about MWGCs such as stellar BH or IMBH content, SNe kicks, and the initial mass function, which affect ejection velocities and the cluster evaporation rate.

\begin{acknowledgements}
This work was supported by NSF grant AST-2108624 and NASA grant 80NSSC22K0722, as well as the computational resources and staff contributions provided for the \texttt{Quest} high-performance computing facility at Northwestern University. \texttt{Quest} is jointly supported by the Office of the Provost, the Office for Research, and Northwestern University Information Technology. G.F. and F.A.R. acknowledge support from NASA grant 80NSSC21K1722. N.C.W acknowledges support from the CIERA Riedel Family Graduate Fellowship. F.K. acknowledges support from a CIERA Board of Visitors Graduate Fellowship. S.C. acknowledges support from the Department of Atomic Energy, Government of India, under project No. 12-R\&D-TFR-5.02-0200 and RTI 4002. K.K. is supported by an NSF Astronomy and Astrophysics Postdoctoral Fellowship under award AST-2001751.
\end{acknowledgements}

\software{\texttt{CMC} \citep{Joshi2000,Joshi2001,Fregeau2003,Fregeau2007,Chatterjee2010,Morscher2013,Pattabiraman2013,CMC_v1.0,CMCRelease}, \texttt{fewbody} \citep{Fregeau2004,Antognini2014,AmaroSeoane2016}, \texttt{SSE}/\texttt{BSE} \citep{Hurley2000,Hurley2002}, \texttt{matplotlib} \citep{Matplotlib}, \texttt{NumPy} \citep{NumPy}, \texttt{SciPy} \citep{SciPy}, \texttt{pandas} \citep{Pandas}, \texttt{Astropy} \citep{Astropy}.}

\bibliography{CMCejecta}
\appendix
\restartappendixnumbering
\section{Escape Criteria and the Tidal Boundary} \label{S:escape_conditions}
Formally defining membership in, and escape from, star clusters is conceptually simple but practically challenging. There are theoretical and modeling nuances that while not discussed above for brevity remain worthy of mention. We begin by reviewing the theoretical energy criterion for escape (Appendix~\ref{S:theoretical_energy_criterion}) before discussing some subtle inconsistencies in the criteria used in cluster modeling (Appendix~\ref{S:code_escape_criteria}). Finally, we include several supplementary tables. Specifically, Table~\ref{table:escape_mechanisms_in_CMC} summarizes the current and planned near-future utilization of escape mechanisms in \texttt{CMC}, Table~\ref{table:escape_causes} lists for each model the cumulative counts of escapers by escape mechanism, and Table~\ref{table:escaper_demographics} lists for each model the cumulative counts of escapers by stellar and binary type.

\subsection{Theoretical Energy Criterion} \label{S:theoretical_energy_criterion}
Star clusters experience a purely static tide in the limit that they circularly orbit within the disk of a radially symmetric galaxy. The motion of objects much less massive than the total cluster mass $M_C$ or galactic mass $M_G$ enclosed by this orbit are the familiar domain of the circular, restricted three-body problem. In this case, the sum of the cluster and galactic potentials---$\phi_c(\vectorbold{r})$ and $\phi_g(\vectorbold{r})$, respectively---is static in the frame rotating with the cluster at angular velocity $\vectorbold{\Omega}=\Omega\hat{z}$, where $\vectorbold{r}$ measures from the center of mass between the cluster and galaxy and $\hat{z}$ is perpendicular to the orbital plane. When neglecting internal gravitational scattering between cluster members, the Jacobi integral $E_{\rm J}$ is conserved for each object in this frame \citep[see][]{BinneyTremaine2008}:
\begin{equation} \label{Eq:EJacobi}
E_{\rm J} = \frac{v^2}{2} + \phi_c(\vectorbold{r}) + \phi_g(\vectorbold{r}) - \frac{\Omega^2}{2}\left(x^2+y^2\right).
\end{equation}
The sum of the last three terms on the right is the effective potential $\phi_{\rm eff}$. Since $v^2>0$, Equation~(\ref{Eq:EJacobi}) implies that objects cannot enter regions where $\phi_{\rm eff} > E_{\rm J}$. The Jacobi--Hill surfaces defined by $\phi_{\rm eff} = E_{\rm J}$ range from fully and separately enclosing the cluster and galactic center (highly negative $E_{\rm J}$) to allowing passage through narrow openings directly toward and away from the galactic center (modestly negative $E_{\rm J}$), to disappearing entirely (high $E_{\rm J}$, excluding no regions).

The highest-$E_{\rm J}$ surface still \textit{fully enclosing} the cluster is the football-shaped Roche surface familiar from binary evolution, whose ends terminate at saddle points of $\phi_{\rm eff}({\vectorbold{r}})$ known as Lagrange points, where $\nabla\phi_{\rm eff}(\vectorbold{r})=0$. So, by definition, objects within the cluster may only cross beyond the Roche surface once they have $E_{\rm J} > \phi_{\rm eff}(r_{\rm J})$, where the Jacobi radius $r_{\rm J}$ is the distance between the cluster and either Lagrange point. Solving $\nabla\phi_{\rm eff}(\vectorbold{r})=0$ in the limit $\mu = M_C/M_G\ll 1$ for a cluster and galaxy that are both point masses results in the expression \citep[e.g.,][]{BinneyTremaine2008} $r_{\rm J}/R_G \approx (\mu/3)^{1/3}$. This result changes negligibly for more realistic cluster potentials, so long as $\nabla \phi_c(r_{\rm J})$ is small, i.e., when the half-mass radius $r_h$ enclosing half the cluster mass satisfies $r_h/r_{\rm J} \ll 1$. However, $r_{\rm J}$ does change significantly for more realistic galactic potentials since much of a galaxy's mass lies beyond the orbit of a typical star cluster ($R_G\sim 10$~kpc in the MW) and so $\nabla \phi_g(r_{\rm J})$ remains steep. A logarithmic MW potential based on the $r^{-2}$ scaling of the MW mass density increases the Jacobi radius to $r_{\rm J}/R_G \approx (\mu/2)^{1/3}$ \citep[e.g.,][]{Spitzer1987}.

Since objects within the cluster may only cross beyond the Roche surface once they have $E_{\rm J} > \phi_{\rm eff}(r_{\rm J})$, $r_{\rm J}$ is a common choice for the definition of a sphericalized tidal radius and the criterion for escape beyond this radius is simply
\begin{equation} \label{Eq:Jacobi_criterion}
E_{\rm J} > \phi_{\rm eff}(r_{\rm J}) \approx -\frac{3}{2} \frac{GM_C}{r_{\rm J}}, 
\end{equation}
where the second equality holds for either definition of $r_{\rm J}$ above---e.g., Equation~(5.6) in \citet{Spitzer1987}. This criterion is necessary but not sufficient for objects to cross beyond $r_{\rm J}$ since $r_{\rm J}$ is the \textit{maximum} distance to the (nonspherical) Roche surface; the \textit{minimum} distance is actually $\approx 2r_{\rm J}/3$ in the $\hat{y}$-direction parallel to the cluster's velocity. As pointed out by \cite{Spitzer1987}, this introduces potential inconsistency between $r_{\rm J}$ and an observational tidal boundary $r_{t,{\rm obs}}$, typically extrapolated from the surface density in regions where the equipotential surfaces are still quite spherical. The observed values of $r_{t,{\rm obs}}$ should in principle include this 2/3 factor: $r_{t,{\rm obs}}\equiv 2r_{\rm J}/3$. Ultimately, the exact definition of an inherently approximate tidal radius is somewhat arbitrary and further alternatives exist, such as a more intermediate value $2r_{\rm J}/3<r_{t,{\rm vol}}<r_{\rm J}$ defined to enclose a spherical volume equal to that enclosed within the actual (nonspherical) Roche surface.

The escape criterion and cluster membership determined by the tidal boundary are nebulous for other reasons, as well. For instance, escape under the above condition is reversible; even once a body's $E_{\rm J}$ is sufficient to allow its escape from the cluster---at which instant it is technically unbound, i.e., a \textit{potential escaper}---gravitational interactions with other bodies on its way out of the cluster can scatter it back down to lower energy. Such back-scattering is a long-known complication to cluster escape \citep[e.g.,][]{Chandrasekhar1942,King1959}. This means that some significant fraction of unbound particles within the Roche surface at any particular instant will become re-bound before escaping beyond it. A negligible but nonzero fraction will even do so after crossing the Roche surface since back-scattering can in principle take place between two different escapers or field stars at arbitrarily large $r$. Even when this does not occur, potential escapers often take significant time (up to gigayears or several relaxation times) to cross beyond the Roche surface \citep[e.g.,][]{FukushigeHeggie2000}, primarily because the escape trajectories may take many crossing times to find and pass through the openings in the Roche surface near the Lagrange points. These openings are very small when the escaper has $E_{\rm J}$ just above $\phi_{\rm eff}(r_{\rm J})$, barely enough to escape. Finally, the trajectories of objects in a rotating frame are complex, made more so by mathematical chaos inherent to the $N$-body problem. The Coriolis effect in particular can bend trajectories back on themselves, allowing semi-stable retrograde orbits about the cluster center to exist beyond $r_{\rm J}$ \citep[e.g.,][]{Henon1970} and past escapers to reenter the cluster at some later time even in the absence of any scattering. Indeed, preliminary results from Weatherford et al. (2023b, in preparation) suggest that the inward flux of previously escaped stars across the spherical surface at $r=r_{\rm J}$ is significant, about ${\sim}10\%$ of the outward flux by a Hubble time. Note this result rests upon the three-body problem of an escaper orbiting in the combined potential of the cluster and its host galaxy, neglecting interactions with field stars, other nearby star clusters, and molecular clouds. (This is reasonable since diffusive perturbations from such sources are generally quite small since galactic halos inhabit the collisionless regime of stellar dynamics.) Similarly, originally external objects may enter the cluster and become bound due to internal gravitational scattering. Except for full trajectories in direct $N$-body models and back-scattering more generally, these effects are commonly neglected in GC models, as much larger uncertainties generally exist in regard to the history of the host environment and its (time-dependent) tidal effects; see Section~\ref{S:tidal_stripping}.

\subsection{Different Escape Criteria in Cluster Modeling} \label{S:code_escape_criteria}
In addition to the complexities in the theoretical escape criterion, there are modeling nuances to be mindful of, too. Few studies with modern cluster modeling codes use the raw theoretical energy criterion in Equation~(\ref{Eq:Jacobi_criterion}) to determine when to remove particles from simulations. This has less to do with the underlying physics and more to do with practicality. Equation~(\ref{Eq:Jacobi_criterion}) is indeed the formally correct criterion for the energy threshold a particle must exceed to \textit{potentially} escape at some future time. In modeling, however, we generally care less about whether a particle is instantaneously bound and more about when we can reasonably remove it from the simulation (happily increasing computational speed) with confidence that doing so will not significantly alter cluster evolution. So modern direct $N$-body codes typically remove particles from simulations only once the particles are far enough away from the cluster center that they are relatively unlikely to return to within the tidal boundary; several times the tidal radius is a common criterion \citep[e.g.,][]{Lee2006,MoyanoLoyolaHurley2013,Rodriguez2016,Kamlah2022}.

Monte Carlo codes also used a radial criterion, removing stars with apocenter $r_a > r_t = r_{\rm J}$, until a modified version of the energy-based criterion was found to better reproduce the escape rate from direct $N$-body results \citep[e.g.,][]{Giersz2008,Chatterjee2010}. The reason is apparent in the observation above that $r_{\rm J}$ is the \textit{maximal} distance to the tidal boundary; stars can satisfy the energy criterion $E_{\rm J} > \phi_{\rm eff}(r_{\rm J})$ with much lower $r_a > (2/3)r_{\rm J}$, so the criterion $r_a > r_t = r_{\rm J}$ dramatically underpredicts the escape rate. The criterion $r_a > 2r_{\rm J}/3$, meanwhile, would overpredict the escape rate as orbits with $(2/3) r_{\rm J} < r_a < r_{\rm J}$ oriented toward/away from the Lagrange points still have $E_{\rm J} < \phi_{\rm eff}(r_{\rm J})$, incapable of escape. However, the energy-based escape criterion requires modification for Monte Carlo codes since they assume a spherical cluster potential. The relevant energy is the orbital energy within only the cluster potential,
\begin{equation} \label{Eq:cluster_energy}
E=v^2/2 + \phi_c(\vectorbold{r}) = E_{\rm J} + \phi_c(\vectorbold{r}) - \phi_{\rm eff}(\vectorbold{r}).
\end{equation}
Expanding for clustercentric distances $r\ll R_G$ yields $\phi_{\rm eff}(\vectorbold{r})\approx \phi_c(\vectorbold{r}) + (\Omega^2/2)(z^2-bx^2)$, where $b=(2,3)$ for logarithmic and point-mass galactic potentials, respectively---see Equation~(5.4) in \citet{Spitzer1987}. From Equation~(\ref{Eq:rtCMC}), $r_{\rm J}^3 \approx M_C R_G^3/(b M_G)$, so by Kepler's third law the square of the cluster's galactocentric angular velocity is $\Omega^2 \approx GM_G/R_G^3 \approx GM_C/(br_{\rm J}^3) = -\phi_c(r_{\rm J})/(br_{\rm J}^2)$. From Eqs.~\ref{Eq:Jacobi_criterion} and \ref{Eq:cluster_energy}, the energy criterion in terms of the orbital energy $E$ within the cluster is then
\begin{equation} \label{Eq:cluster_energy_criterion}
E > \beta\phi_c(r_{\rm J});\ \ \ \ \beta\equiv3/2-(bx^2-z^2)/(2br_{\rm J}^2) \approx 3/2.
\end{equation}
This criterion can be fully sphericalized by projecting $\vectorbold{r}/r_{\rm J}=(x,y,z)/r_{\rm J}$ randomly onto the unit sphere to obtain $\langle\beta\rangle\equiv 3/2 - (r/r_{\rm J})^2/12$. However, since most escapers originate deep within the cluster ($r/r_{\rm J}\ll 1$), the second term on the right can be neglected as in Equation~(\ref{Eq:cluster_energy_criterion}). The \cite{Giersz2008} energy criterion---Equation~(\ref{Eq:Giersz_energy_criterion})---used by \texttt{CMC} modifies this by adding a term making escape slightly harder to account for back-scattering. This back-scattering term decreases with $N$ since individual particles have a less dominant influence on each other in more populous clusters. The \texttt{MOCCA} Monte Carlo code currently implements an even further-updated escape criterion \citep[][]{Giersz2013} that eliminates the \cite{Giersz2008} criterion's $N$-dependence, which was only validated against direct $N$-body models for relatively small $N$. This update also introduces a lag time to allow particles meeting the energy criterion time to escape the cluster before removing them from the simulation, but has minimal impact on the cluster escape rate \citep[see][]{Giersz2013} and so has not been incorporated into \texttt{CMC}. This is of little consequence to the analysis in later papers in this series, since they directly solve for the time to escape by integrating escaper orbits from the point of ejection to beyond the tidal boundary in the full potential of the cluster and Galaxy.

Finally, note that while the energy-based criterion in \texttt{CMC} best matches direct $N$-body results for clusters with static tides, the apocenter criterion may work better for cluster models with time-dependent tides from an elliptical orbit within their host galaxy \citep[e.g.,][]{Rodriguez2022a}. More complex distance-based criteria that better reflect the shape of the tidal boundary may also improve comparison to direct $N$-body models \citep[e.g.,][]{Sollima2014}. These concerns, and tidal physics in the Monte Carlo method more generally, merit further investigation.

\begin{deluxetable*}{m{0.08\textwidth} | m{0.15\textwidth} m{0.19\textwidth} m{0.10\textwidth} m{0.38\textwidth}}
\centering
\tabletypesize{\footnotesize}
\tablewidth{0pt}
\tablecaption{List of Escape Mechanisms and Status in \texttt{CMC}}
\label{table:escape_mechanisms_in_CMC}
\tablehead{
    \colhead{} &
	\colhead{Mechanism Type} &
	\colhead{Mechanism} &
	\colhead{Implementation} &
	\colhead{Status in \texttt{CMC}}
}
\startdata
\multirow[c]{5}{*}{Evaporation}
& \multirow[c]{2}{\linewidth}{Two-body relaxation with static tides} & Two-body relaxation   & \centering ACTIVE &
Toggleable ON/OFF; adjustable Coulomb logarithm (default 0.01) and max. deflection angle (default $\sqrt{2}$).\\
&                                                                 & Static galactic tide     & \centering ACTIVE &
Toggleable ON/OFF; circular cluster orbit in logarithmic galactic potential with adjustable circular velocity ($220{\rm\ km\ s}^{-1}$). \\ \cline{2-5}
& \multirow[c]{3}{\linewidth}{Time-dependent tides}               & Elliptical/inclined orbit & \centering OPTIONAL   &
Partially toggleable ON/OFF by providing at simulation start a data table describing a time-varying tidal tensor. \\
&                                                                 & Tidal shocking           & \centering  ---   & Implementation under consideration, but is challenging since \texttt{CMC} operates on the relaxation timescale while shocks can operate on the crossing timescale. \\
\hline \hline \multirow[c]{12}{\linewidth}{Ejection}
& \multirow[c]{6}{\linewidth}{Strong encounters} & Strong two-body encounters  & \centering   ---   & Implementation with \texttt{fewbody} likely by 2024. \\
&                                             & Three-body binary formation        & \centering ACTIVE  &
Toggleable ON/OFF (for all bodies, just BHs, or none at all) semi-analytic prescription \citep[][]{Morscher2013}; can set minimum hardness ratio (default $\eta = 1$). Upgraded implementation with \texttt{fewbody} possible by 2024. \\
&                                             & Binary--single                  & \centering ACTIVE  & Toggleable ON/OFF. \\
&                                             & Binary--binary                  & \centering ACTIVE  & Toggleable ON/OFF. \\
&                                             & Higher-order encounters        & \centering   ---   & Implementation possible but largely unwarranted (probably only relevant in very dense young clusters). \\
&                                             & Unstable triple disintegration & \centering PARTIAL &
Not toggleable; \texttt{CMC} breaks apart and records all formed triples at birth. \\ \cline{2-5}
& \multirow{3}{\linewidth}{(Near-)contact recoil}       & Direct physical collisions   & \centering PARTIAL  &
Toggleable ON/OFF; idealized sticky-sphere prescription without asymmetric mass loss makes this not yet a true ejection mechanism. \\
&                                                       & TDEs      & \centering   ---   &
None so far; while TDE rates have been studied with \texttt{CMC} in post-processing, prescriptions to modify stars as a TDE occurs are under active development. \\
&                                                       & GW-driven mergers   & \centering ACTIVE  &
Toggleable ON/OFF; post-Newtonian terms included in \texttt{CMC}'s \texttt{fewbody} integrator. \\ \cline{2-5}
& \multirow[c]{3}{\linewidth}{Stellar evolution recoil} & BH SNe kicks   & \centering ACTIVE &
Adjustable prescription; see text and/or \citet{CMCCatalog}. \\
&                                                       & NS SNe kicks & \centering ACTIVE &
Adjustable prescription; see text and/or \citet{CMCCatalog}. \\
&                                                       & WD kicks            & \centering  ---   &
Not currently implemented, though uniform $2--9$~km/s kicks were tested in a past \texttt{CMC} version \citep{Fregeau2009}.
\enddata
\end{deluxetable*}

\begin{deluxetable*}{l|ccccccccccccc}
\tabletypesize{\footnotesize}
\tablewidth{0pt}
\tablecaption{Cumulative Counts of Escapers by Type and Escape Mechanism}
\label{table:escape_causes}
\tablehead{
 \colhead{} & \multicolumn{7}{c}{Escapers with BH(s)} & \colhead{With NS(s)} & \multicolumn{5}{c}{Without BHs or NSs}  \\[-0.2cm]
 \colhead{} & \colhead{BBH Merger} & \colhead{Own SN} & \colhead{Comp. SN} & \colhead{3BB} & \colhead{BS} & \colhead{BB} &
 \colhead{Relax.} & \colhead{} & \colhead{Comp. SN} & \colhead{3BB} & \colhead{BS} & \colhead{BB} & \colhead{Relax.} \\[-0.2cm]
 \colhead{} & \colhead{(1)} & \colhead{(2)} & \colhead{(3)} & \colhead{(4)} & \colhead{(5)} & \colhead{(6)} &
 \colhead{(7)} & \colhead{(8)} & \colhead{(9)} & \colhead{(10)} & \colhead{(11)} & \colhead{(12)} & \colhead{(13)}
}
\startdata
 1 & 29 & 417 & 5 & 1 & 640 & 183 & 12 & 4945 & 25 &   9753 & 11,783 & 1551 & 397,922 \\
 2 & 27 & 401 & 2 & 2 & 628 & 178 & 14 & 4806 & 35 & 10,217 & 11,293 & 1561 & 172,686 \\
 3 & 35 & 418 & 3 & 0 & 628 & 189 & 10 & 4732 & 22 & 10,763 &   9225 & 1216 & 110,691 \\ \hline
 4 & 33 & 498 & 7 & 0 & 625 & 171 & 13 & 4872 & 50 & 12,625 &   3397 &  427 & 286,870 \\
 5 & 30 & 487 & 4 & 3 & 604 & 208 &  9 & 4812 & 44 & 13,598 &   3234 &  444 & 101,547 \\
 6 & 34 & 486 & 5 & 4 & 601 & 194 &  8 & 4829 & 48 & 13,266 &   3450 &  432 &  58,592 \\ \hline
 7 & 17 & 532 & 2 & 1 & 540 & 177 & 20 & 4998 & 84 & 14,142 &   3216 &  439 & 328,236 \\
 8 & 19 & 528 & 2 & 3 & 572 & 148 & 14 & 4943 & 81 & 13,459 &   2914 &  377 &  65,910 \\
 9 & 15 & 530 & 4 & 2 & 527 & 165 & 12 & 4955 & 78 & 14,031 &   3026 &  436 &  40,859 \\ \hline
10*&  5 & 560 & 6 & 5 & 337 & 137 & 56 & 5101 & 93 & 10,568 &   3588 &  605 & 772,543 \\
11 & 10 & 550 & 5 & 6 & 402 & 152 & 14 & 5026 & 93 & 14,162 &   3461 &  581 &  59,361 \\
12 & 14 & 551 & 9 & 4 & 392 & 136 & 13 & 5007 & 92 & 12,370 &   2867 &  450 &  28,036 \\
\enddata
\tablecomments{\footnotesize The number of escapers (singles or binaries) containing at least one BH (1-7), at least one NS but no BH (8), and everything else (neither NSs nor BHs; 9-13). The first and third of these categories are further divided by escape mechanism: a binary BH merger kick (1), the escaper's own SN (2), a (former) binary companion's SN (3), three-body binary formation (4,10), a binary--single \texttt{fewbody} encounter (5,11), a binary--binary \texttt{fewbody} encounter (6,12), and two-body relaxation (7,13). The * on simulation 10 indicates that the values are not entirely reliable since the cluster fully disrupted and the final totals therefore include late-stage evolution on a dynamical timescale, which \texttt{CMC} cannot accurately capture.}
\end{deluxetable*}

\begin{deluxetable*}{l|cccccccccccccccc}
\tabletypesize{\scriptsize}
\tablewidth{0pt}
\tablecaption{Escaper Demographics: Cumulative Population Counts and Binarity}
\label{table:escaper_demographics}
\tablehead{
	\colhead{} &
	\colhead{$f_b$~(\%)} &
	\colhead{Total} &
	\colhead{MS} &
	\colhead{G} &
	\colhead{WD} &
	\colhead{NS} &
	\colhead{BH} &
	\colhead{BHBH} &
	\colhead{BHNS} &
    \colhead{BHWD} &
	\colhead{NSNS} &
	\colhead{NSWD} &
	\colhead{WDWD} &
    \colhead{BHMS} &
	\colhead{NSMS} &
	\colhead{WDMS} \\[-0.2cm]
	\colhead{} & \colhead{(1)} & \colhead{(2)} & \colhead{(3)} & \colhead{(4)} & \colhead{(5)} & \colhead{(6)} & \colhead{(7)} & \colhead{(8)}
	& \colhead{(9)} & \colhead{(10)} & \colhead{(11)} & \colhead{(12)} & \colhead{(13)} & \colhead{(14)} & \colhead{(15)} & \colhead{(16)}
}
\startdata
 1 & 4.1 & 427,266 & 420,384 & 348 & 17,664 & 4949 & 1412 & 125 & 2 & 0 & 2 & 28 & 59 & 29 & 15 & 114 \\
 2 & 4.0 & 201,850 & 197,457 & 161 &   6111 & 4807 & 1374 & 122 & 1 & 2 & 0 & 18 & 34 & 11 & 10 &  73 \\
 3 & 3.8 & 137,932 & 132,965 & 118 &   3928 & 4732 & 1398 & 115 & 0 & 2 & 0 &  3 & 24 & 14 &  1 &  42 \\ \hline
 4 & 3.9 & 309,588 & 303,100 & 314 & 11,851 & 4872 & 1461 & 114 & 0 & 0 & 0 &  3 &  1 &  9 &  1 &  17 \\
 5 & 3.5 & 125,024 & 118,880 & 147 &   4111 & 4812 & 1457 & 112 & 0 & 1 & 0 &  1 &  0 &  3 &  0 &   8 \\
 6 & 3.3 &  81,949 &  75,756 &  97 &   2516 & 4829 & 1454 & 122 & 0 & 0 & 0 &  1 &  1 &  9 &  0 &   9 \\ \hline
 7 & 4.1 & 352,404 & 342,152 & 456 & 17,629 & 4999 & 1421 & 132 & 0 & 0 & 1 &  8 &  2 & 11 &  1 &  30 \\
 8 & 3.4 &  88,970 &  81,851 & 107 &   3647 & 4943 & 1406 & 120 & 0 & 0 & 0 & 12 &  0 &  2 &  2 &  11 \\
 9 & 3.2 &  64,640 &  57,525 &  85 &   2720 & 4955 & 1381 & 126 & 0 & 0 & 0 &  4 &  0 &  7 &  1 &  10 \\ \hline
10 & 4.7 & \multicolumn{15}{c}{\textit{Disrupted}} \\
11 & 3.5 &  83,823 &  76,071 & 131 &   4276 & 5026 & 1238 &  99 & 0 & 0 & 0 & 15 &  3 &  4 &  1 &  37 \\
12 & 3.0 &  49,941 &  42,979 &  71 &   2163 & 5007 & 1221 & 102 & 0 & 0 & 0 & 10 &  1 &  5 &  1 &  26 \\
\enddata
\tablecomments{\footnotesize The cumulative binary fraction $f_{\rm b}$ (1) and cumulative number of escapers by type (2-16). Columns 2-7 each list the total number of escaped objects (separately counting the primary and secondary in the case of a binary). Columns 8-16 each list the number of escaped binaries of different types. Throughout all columns, MS = main-sequence star, G = giant (including Hertzsprung gap objects), WD = white dwarf, NS = neutron star, and BH = black hole.}
\end{deluxetable*}

\end{document}